\documentclass{article}
\usepackage{romp}

\usepackage[dvips]{graphicx}
\usepackage{amsmath}
\usepackage{amsfonts}
\usepackage{amssymb}
\usepackage{longtable} 
\usepackage{color}

\allowdisplaybreaks[4] 

\newcommand{\df}{\ {\overset {\rm def} =}\ }
\newcommand{\dr}[2]{\frac {{\rm d} {#1}} {{\rm d} {#2}}}

\newcommand{\dril}[2]{{{\rm d} {#1}} / {{\rm d} {#2}}}
\newcommand{\pdril}[2]{{\partial {#1}} / {\partial {#2}}}

\newcommand{\llim}[1] {\ {\underset {#1} {\longrightarrow}}\ }

\title{Causality in the maximally extended Reissner--Nordstr\"{o}m spacetime
with identifications}
\author{Andrzej Krasi\'nski \\ N. Copernicus Astronomical Centre, Polish Academy
of Sciences \\
Bartycka 18, 00 716 Warszawa, Poland \\
e-mail: akr@camk.edu.pl}

\begin{document}

\maketitle
\begin{abstract}
The maximally extended Reissner--Nordstr\"{o}m (RN) spacetime with $e^2 < m^2$
can be interpreted either as an infinite chain of asymptotically flat regions
connected by tunnels between timelike singularities or as a set of just one pair
of asymptotically flat regions and one tunnel; the repetitions of this set in
the infinite chain being identified. The second interpretation gives rise to the
suspicion of acausality, i.e. the possibility of sending messages to one's own
past. A numerical investigation of this problem was carried out in this paper
and gave the following result. Let E be the initial point of a radial timelike
future-directed ingoing geodesic G, lying halfway between the outer horizon and
the image of the null infinity in the maximally extended RN spacetime. Let E$'$
be the first future copy of E. It was verified whether the turning point of G
lies outside or inside the past light cone (PLC) of E$'$. In the second case the
breach of causality does occur. It turned out that the acausality is present
when $V_E$, the timelike coordinate of E, is negative with a sufficiently large
$|V_E|$, and is absent with a sufficiently large $V_E > 0$. In between these
values there exists a $\widetilde{V}_E$, dependent on the initial data for the
geodesic, for which the turning point lies on the PLC. So, the identification
does lead to acausality. Nonradial timelike and null geodesics were also
investigated, and a few hitherto unknown properties of the maximal extension
were revealed. For example, the singularity arc at $r = 0$ may be convex or
concave, depending on the values of $m$ and $e$.
\end{abstract}

\section{Motivation and summary}\label{intro}

The maximally extended Reissner \cite{Reis1916} -- Nordstr\"{o}m \cite{Nord1918}
(RN) spacetime with $e^2 < m^2$ can be interpreted either as an infinite chain
of asymptotically flat regions connected by tunnels between timelike
singularities or as a set of just one pair of asymptotically flat regions and
one tunnel; the repetitions of this set in the infinite chain being identified.
The identification may be suspected of leading to acausality (i.e. an observer
could supposedly send a message to its own past by means of timelike or null
geodesics). A radial {\it null} geodesic sent into the tunnel will hit the
singularity and will not get out into the next asymptotically flat region unless
it is reflected somewhere in the tunnel. The problem was to verify what happens
with timelike geodesics, radial and nonradial, and with nonradial null
geodesics.

In the present paper it was shown by numerical examples that the breach of
causality does or does not occur depending on the initial point of the timelike
geodesic. The method of computation was as follows. We first numerically
integrated a future-directed ingoing radial timelike geodesic $G_1$ emitted
outside the outer RN horizon $r = r_+ \df m + \sqrt{m^2 - e^2}$ for which the
radial coordinate of the emission event was at a midpoint between the image of
$r = r_+$ and the image of the null infinity, and the emission time was
distinctly later than the instant of time-symmetry. In the $(u, v)$ coordinates
adapted to the $r_+$ horizon, $G_1$ smoothly crossed $r = r_+$. At a point $P_1$
in the region $r_- < r < r_+$ (where $r_- \df m - \sqrt{m^2 - e^2}$) the
coordinates were transformed to the $(u', v')$ adapted to the $r_-$ horizon, and
the numerical integration was continued from point $P_3$ (the image of $P_1$
under this transformation) smoothly through $r = r_-$. The geodesic $G_1$ was
followed up to its turning point (TP) denoted $P_5$, where, at $r = r_{\rm tp}$,
it would change to an outgoing one. Then, we issued from event E$'$ -- the first
future copy of E that would coincide with E under the identification -- a
past-directed ingoing radial {\it null} geodesic $G_2$. The $G_2$ is the radial
generator of the past light cone (PLC) of E$'$. The equation of a radial null
geodesic can be solved explicitly. It was followed through the $r_+$ horizon,
and in the region $r_- < r < r_+$ the coordinates were changed from $(u, v)$ to
$(u', v')$. In the new coordinates $G_2$ was continued until it reached $r =
r_{\rm tp}$. There, it turned out that $P_5$ lies to the future of the PLC.
Hence, if $G_1$ (the first geodesic) were continued to the future of $P_5$, it
would not enter the PLC of E$'$ and near E$'$ would lie to the future of E$'$.
Consequently, the observer at E would not be able to send any message to its
past. Nonradial timelike and null geodesics behave similarly: the $(u', v')$
coordinates of their TPs differ from those of $P_5$, but not sufficiently to
change the conclusion.

However, if the initial point E lies at the same $r$ as before, but distinctly
earlier than the instant of time-symmetry, then the opposite occurs: the TP of a
radial timelike ingoing future-directed geodesic lies to the past of the PLC of
E$'$ and sending a message to one's own past is possible. A logical conclusion
is that if E lies somewhere between the two previously mentioned locations, then
the TP will lie right on the PLC of E$'$, which was also verified in this paper.
The exact location of this preferred E depends on the parameters of $G_1$.

The final conclusion is that the identification does lead to acausality.

The paper is organised as follows. In Sec. \ref{RNintro}, the basic geometric
properties of the RN spacetime with $e^2 < m^2$ are described. The presentation
follows the reasoning of Graves and Brill \cite{GrBr1960} (GB) with several
extensions. Ref. \cite{GrBr1960} presented the main idea on how to remove the
spurious singularities, but left some details and consequences for the readers
to fill in. For numerical computations everything must be stated explicitly. In
particular, the $(u, v)$ coordinates that remove the spurious singularities at
the horizons and their relation to the $(U, V)$ coordinates used in the
conformal diagrams are discussed in detail. The maximal extension of the RN
metric is re-derived by the GB method. A surprise emerges: the arcs of the
singularity at $r = 0$ are concave or convex depending on the values of $m$ and
$e$. The graphs of the maximal extension shown in Refs. \cite{GrBr1960} and
\cite{PlKr2024} are correct only for the distance between the vertices of the
hyperbola $u^2 - v^2 =$ constant being sufficiently small. Also, the
transformation from the $(t, r)$ to the $(u, v)$ coordinates in Ref.
\cite{GrBr1960} covers only one of four sectors of the $(u, v)$ coordinate
plane. The sectors are separated by the $u = \pm v$ straight lines, and the
transformation is different in each sector.

In Sec. \ref{geode}, the geodesic equations in the $(u, v)$ coordinates are
derived and discussed. They turn out to be the same in each of the four sectors.

In Sec. \ref{thepro}, the transformations between the $(u, v)$ coordinates
adapted to $r_+$ and the $(u', v')$ adapted to $r_-$ are derived and discussed.
It is shown that the geodesic equations in the $(u', v')$ coordinates are
identical to those in the $(u, v)$ coordinates.

In Sec. \ref{shaper}, the method of identifying the lines of constant $r$, in
particular of the locus of TPs of radial timelike geodesics, is explained and
discussed. The transformations involved in constructing the maximal extension
are illustrated by numerical examples.

In Secs. \ref{numcal} and \ref{PLCE}, the numerical integration of radial
timelike and null geodesics is explained step by step and it is proved that a
radial timelike geodesic emitted sufficiently late cannot enter the past light
cone of the first future copy of the emitter.

In Secs. \ref{norad} and \ref{nunorad} the same is shown (numerically) for a
late-emitted nonradial timelike geodesic with the absolute value of the angular
momentum constant $J_0$ being near the allowed maximum, and for a nonradial null
geodesic with the same $J_0$. (Geodesics with larger $|J_0|$ do not enter the $r
< r_+$ region, so they cannot propagate through the tunnel between the
singularities and are irrelevant for the problem of causality).

In Sec. \ref{newsec}, it is demonstrated that an early-emitted ingoing radial
timelike geodesic has its TP earlier than the past light cone of E$'$, so does
lead to acausality. It is also demonstrated that there exists a timelike ingoing
radial geodesic emitted at a time between the early and the late one, for which
the TP lies right on (E$'$)'s past light cone.

In Sec. \ref{conclu} the conclusions and implications of the results of this
paper are summarised and discussed. In particular, the geometrical peculiarities
of the maximal extension that are not visible at the level of a general
discussion, but clearly appear in the numerical computations, are pointed out.

Some details of the calculations are explained in five appendices.

\section{Basic facts about the maximally extended Reissner--Nordstr\"{o}m
spacetime}\label{RNintro}

\setcounter{equation}{0}

The signature $(+ - - -)$ will be used throughout the paper.

The RN metric is the electrovacuum solution of the Einstein--Maxwell equations
that describes the spacetime in a neighbourhood of a spherically symmetric body
(or a black hole) of mass $m$ and electric charge $e$. In curvature coordinates
it is
\begin{equation}\label{2.1}
{\rm d} s^2 = \phi {\rm d} t^2 - (1 / \phi) {\rm d} r^2 - r^2 \left({\rm d}
\vartheta^2 + \sin^2 \vartheta {\rm d} \varphi^2\right),
\end{equation}
where
\begin{equation}\label{2.2}
\phi = 1 - 2m / r + e^2 / r^2.
\end{equation}
The mass $m$ and the charge $e$ are expressed in units of length. They are
related to the mass $M$ and charge $Q$ in physical units by $m = GM/c^2$ and $e
= \sqrt{G} Q / c^2$, where $G$ is the gravitational constant and $c$ is the
velocity of light (see Eq. (19.62) in Ref. \cite{PlKr2024}).

We shall consider only those RN metrics for which $e^2 < m^2$. When $e^2 > m^2$,
the spurious singularities do not exist (only the genuine one at $r = 0$ is
there) so no extension is needed, and the case $e^2 = m^2$ is less challenging
mathematically. The metric with $e^2 < m^2$ has spurious singularities (event
horizons) where $\phi = 0$, i.e. at
\begin{equation}\label{2.3}
r_- = m - \sqrt{m^2 - e^2}, \qquad r_+ = m + \sqrt{m^2 - e^2}.
\end{equation}
In the Schwarz\-schild limit $e \to 0$, the horizon at $r = r_-$ collapses on
the genuine singularity at $r = 0$, while the other one goes over into the
Schwarzschild horizon at $r = 2m$. The sets (\ref{2.3}) are made nonsingular by
coordinate transformations as follows \cite{GrBr1960}.

We introduce such coordinates $u(t, r)$ and $v(t, r)$ in which
\begin{equation}\label{2.4}
{\rm d} s^2 = f^2(u, v) \left({\rm d} v^2 - {\rm d} u^2\right) - r^2(u, v)
\left({\rm d} \vartheta^2 + \sin^2 \vartheta {\rm d} \varphi^2\right).
\end{equation}
The functions $f$, $u$ and $v$ then obey
\begin{eqnarray}\label{2.5}
f^2 \left({v,_t}^2 - {u,_t}^2\right) &=& \phi(r), \qquad f^2 \left({v,_r}^2 -
{u,_r}^2\right) = - \frac 1 {\phi(r)}, \nonumber \\
v,_t v,_r - u,_t u,_r &=& 0.
\end{eqnarray}
This implies ${u,_t}^2/{v,_r}^2 = \phi^2(r)$. From this and from the last of
(\ref{2.5}) we obtain\footnote{Formally, there exists the second solution $u,_t
= - \phi v,_r$, $v,_t = - \phi u,_r$, but it is equivalent to (\ref{2.6}) by the
coordinate transformation $t = -t'$, which is an isometry of (\ref{2.1}) --
(\ref{2.2}).}
\begin{equation}\label{2.6}
u,_t = \phi(r) v,_r, \qquad v,_t = \phi(r) u,_r.
\end{equation}
We now introduce the new variable $r^*(r)$ by
\begin{equation}\label{2.7}
\dril {r^*} r = 1/ \phi,
\end{equation}
then the solution of (\ref{2.6}) is
\begin{equation}\label{2.8}
u = h(r^* + t) + g(r^* - t), \qquad v = h(r^* + t) - g(r^* - t),
\end{equation}
where $h$ and $g$ are arbitrary functions (an additive constant in $v$ has been
ignored because it does not enter (\ref{2.5})). Primes will denote the
derivatives of $h$ and $g$ by their arguments. With $\phi$ given by (\ref{2.2}),
the explicit formula for $r^*(r)$ is
\begin{equation}
r^* = r + \frac {{r_+}^2} {r_+ - r_-} \ln \left|r - r_+\right| - \frac {{r_-}^2}
{r_+ - r_-} \ln \left|r - r_-\right|. \label{2.9}
\end{equation}
Using (\ref{2.8}), we find from (\ref{2.5})
\begin{equation}\label{2.10}
f^2 = \frac {\phi(r)} {4h'(r^* + t) g'(r^* - t)}.
\end{equation}
Any zero of $\phi(r)$ must now be cancelled by the denominator, and the
resulting $f$ must be time-independent, so that (\ref{2.4}) is static and
nonsingular at $\phi = 0$.

The product $h'(r^* + t) g'(r^* - t)$ will be independent of $t$ only if
\begin{equation}\label{2.11}
h = A {\rm e}^{\gamma(r^* + t)} + C, \qquad g = B {\rm e}^{\gamma(r^* - t)} + D,
\end{equation}
where $A$, $B$, $C$, $D$ and $\gamma$ are arbitrary constants. We shall take $C
= D = 0$ (they do not appear in (\ref{2.10})) and, for the beginning, $0 < A =
B$. Then (\ref{2.10}) becomes
\begin{equation}\label{2.12}
f^2 = \frac {\phi(r)} {4 A^2 \gamma^2 {\rm e}^{2 \gamma r^*}}.
\end{equation}
Substituting (\ref{2.11}) in (\ref{2.8}) we obtain the formulae for the
transformation $(t, r) \to (u, v)$:
\begin{eqnarray}\label{2.13}
u &=& A {\rm e}^{\gamma r^*} \left({\rm e}^{\gamma t} + {\rm e}^{- \gamma
t}\right) \equiv 2 A {\rm e}^{\gamma r^*} \cosh (\gamma t), \nonumber \\
v &=& A {\rm e}^{\gamma r^*} \left({\rm e}^{\gamma t} - {\rm e}^{- \gamma
t}\right) \equiv 2 A {\rm e}^{\gamma r^*} \sinh (\gamma t).
\end{eqnarray}
This implies $u \geq 0$ and $|v| \leq u$ ($|v| \to u$ at $t \to \pm \infty$ and
at ${\rm e}^{\gamma r^*} \to 0$), i.e., (\ref{2.13}) covers only the $u \geq 0$
part\footnote{As (\ref{2.9}) shows, $r^* \to -\infty$ when $r \to r_+$ and $r^*
\to + \infty$ when $r \to r_-$, so, depending on the sign of $\gamma$, ${\rm
e}^{\gamma r^*} \to 0$ at $r \to r_+$ or at $r \to r_-$, hence $u \geq 0$ rather
than $u > 0$.} of the region where $u^2 - v^2 \geq 0$. The inverse
transformation $(u, v) \to (t, r)$ is implicitly given by
\begin{equation}\label{2.14}
4 A^2 {\rm e}^{2 \gamma r^*} = u^2 - v^2, \qquad t = \frac 1 {\gamma} {\rm
artanh} (v/u) \equiv \frac 1 {2 \gamma}\ \ln \frac {1 + v / u} {1 - v / u}.
\end{equation}
Thus, in the $(u, v)$ coordinate plane, $t =$ constant on straight lines through
the origin and $r =$ constant on the hyperbolae $u^2 - v^2 =$ constant. From
(\ref{2.12}) and (\ref{2.14}),
\begin{equation}\label{2.15}
f^2 = \frac {\phi} {\gamma^2 \left(u^2 - v^2\right)}.
\end{equation}

To extend the transformation $(t, r) \to (u, v)$ to the $u^2 - v^2 < 0$ region
we choose $C = D = 0$ and $0 > B = -A$ in (\ref{2.11}), and then (\ref{2.13}) --
(\ref{2.15}) are replaced by
\begin{eqnarray}
u = 2 A {\rm e}^{\gamma r^*} \sinh (\gamma t), &\qquad& v = 2 A {\rm e}^{\gamma
r^*} \cosh (\gamma t), \label{2.16} \\
4 A^2 {\rm e}^{2 \gamma r^*} = v^2 - u^2, &\qquad& t = \frac 1 {\gamma} {\rm
artanh} (u/v), \label{2.17}
\end{eqnarray}
the function $f^2$ being still given by (\ref{2.15}).

\begin{figure}[h]
 \begin{center}
 ${}$ \\[-3cm]
 \includegraphics[scale=0.6]{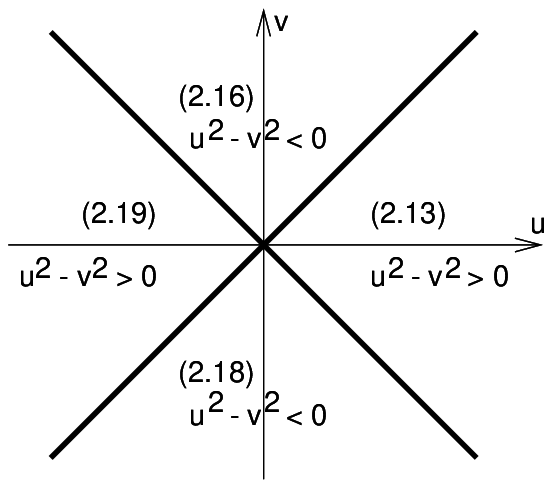}
 \caption{
 \label{sectors}
 \footnotesize
The sectors of the $(u, v)$ coordinate plane covered by different $(t, r) \to
(u, v)$ tansformations. See the text for details.}
 \end{center}
 \end{figure}

The two $(t, r) \to (u, v)$ transformations given by (\ref{2.13}) and
(\ref{2.16}) cover only the $v > - u$ half of the $(u, v)$ coordinate plane, see
Fig. \ref{sectors}. For $\{v < 0, u^2 - v^2 < 0\}$ the covering is provided by
\begin{equation}
u = 2 A {\rm e}^{\gamma r^*} \sinh (\gamma t), \qquad v = - 2 A {\rm e}^{\gamma
r^*} \cosh (\gamma t), \qquad A > 0. \label{2.18}
\end{equation}
Finally, for $\{u < 0, u^2 - v^2 > 0\}$ the covering is provided by
\begin{equation}
u = - 2 A {\rm e}^{\gamma r^*} \cosh (\gamma t), \qquad v = 2 A {\rm e}^{\gamma
r^*} \sinh (\gamma t), \qquad A > 0. \label{2.19}
\end{equation}

When $e^2 < m^2$, Eq. (\ref{2.2}) can be written as
\begin{equation}\label{2.20}
\phi = \frac {(r - r_+) (r - r_-)} {r^2},
\end{equation}
and the expression for ${\rm e}^{2 \gamma r^*}$ is, using (\ref{2.9}),
\begin{equation}
{\rm e}^{2 \gamma r^*} = {\rm e}^{2 \gamma r} \left|r - r_+\right|^{2 \gamma
{r_+}^2 / (r_+ - r_-)} \left|r - r_-\right|^{-2 \gamma {r_-}^2 / (r_+ - r_-)}.
\label{2.21}
\end{equation}
The function ${\rm e}^{2 \gamma r^*}$ becomes infinite or zero at $r = r_+$ and
$r = r_-$, depending on the sign of $\gamma$ (see below). Using (\ref{2.20}) and
(\ref{2.21}), Eq. (\ref{2.12}) becomes
\begin{equation}\label{2.22}
f^2 = \frac {(r - r_+) (r - r_-) \left|r - r_+\right|^{- 2 \gamma {r_+}^2/(r_+ -
r_-)} \left|r - r_-\right|^{2 \gamma {r_-}^2/(r_+ - r_-)}} {4 A^2 \gamma^2 r^2
{\rm e}^{2\gamma r}}.
\end{equation}
Now $\gamma$ can be chosen so as to cancel one of the spurious singularities,
but not both at once. Let $r_1 = r_+$ and $r_2 = r_-$. To cancel the singularity
at $r_i$ we take
\begin{equation}\label{2.23}
\gamma_i = \frac {r_i - r_j} {2{r_i}^2}, \qquad i \neq j.
\end{equation}
The transformations that cancel the singularities at $r = r_+$ and $r = r_-$
will be called, respectively, the $\gamma_1$ and $\gamma_2$ transformation. With
$\gamma = \gamma_1$ and $r > r_+$, Eq. (\ref{2.22}) becomes
\begin{equation}\label{2.24}
f^2 = \left(r - r_-\right)^{1 + {r_-}^2/{r_+}^2} / \left(4 A^2 {\gamma_1}^2 r^2
{\rm e}^{2 \gamma_1 r}\right).
\end{equation}
For this $f^2$ the metric (\ref{2.4}) can be smoothly continued through $r =
r_+$, but $f^2 = 0$ at $r = r_-$, so $r = r_-$ is still a singularity.

With $\gamma = \gamma_2 = \left(r_- - r_+\right) / (2 {r_-}^2)$ and $r < r_+$,
we obtain for $f^2$:
\begin{equation}\label{2.25}
f^2 = {\rm sign}(r_- - r) \left(r_+ - r\right)^{1 + {r_+}^2/{r_-}^2} / \left(4
A^2 {\gamma_2}^2 r^2 {\rm e}^{2 \gamma_2 r}\right).
\end{equation}
This is zero at $r = r_+$ and well-defined for all $r < r_+$ (since this $f^2 <
0$ for $r_- < r < r_+$, Eq. (\ref{2.4}) shows that in this range $u$ is the
time- and $v$ is the space-coordinate). With such $f^2$, the metric can be
continued through $r = r_-$. For numerical computation, a more detailed analysis
of (\ref{2.21}) is needed -- see Appendix \ref{anal219}.

{}From (\ref{2.21}) ${\rm e}^{2 \gamma r^*}(r_i) = 0$ when $\gamma = \gamma_i$.
Thus, from (\ref{2.14}), the set $r = r_i$, in the coordinates that make it
nonsingular, has the equation $u = \pm v$. Again by (\ref{2.21}), ${\rm e}^{2
\gamma r^*}(0) =$ constant $> 0$. Hence, in the part of the $(u, v)$ coordinate
plane where (\ref{2.14}) applies, the equation of the singular set $r = 0$ is
$u^2 - v^2 =$ constant $> 0$, i.e., it is a pair of hyperbolae that intersect
the $u$-axis and are convex towards smaller $|u|$. However, where (\ref{2.16})
applies, the constant $u^2 - v^2$ is negative, and the image of the singular set
is a pair of hyperbolae that intersect the $v$ axis. See Sec. \ref{shaper} for
more on this.

It is useful to map the $(u, v)$ surface into a finite set in such a way that
null geodesics are mapped into themselves (this is called conformal, or Penrose,
mapping). The radial null geodesics in (\ref{2.4}) are $u \pm v =$ constant. We
first introduce the null coordinates
\begin{equation}\label{2.26}
p = u + v, \qquad q = u - v,
\end{equation}
and then define
\begin{equation}\label{2.27}
P = \tanh p, \qquad Q = \tanh q.
\end{equation}
In the $(P, Q)$ coordinates, the equation of the horizon at $u = v$ is $Q = 0$,
and of the one at $u = - v$ it is $P = 0$. The image of the whole $(u, v)$
surface fits in the square $(P, Q) \in \{[-1, 1] \times [-1, 1]\}$, and the null
infinities $p = \pm \infty$, $q = \pm \infty$ are mapped into the sets $P = \pm
1$, $Q = \pm 1$. We introduce the time-space coordinates in this square by
\begin{equation}\label{2.28}
U = (P + Q)/2, \qquad V = (P - Q)/2.
\end{equation}
The transformation between the $(u, v)$ of (\ref{2.12}) -- (\ref{2.19}) and the
$(U, V)$ of (\ref{2.28}) is
\begin{eqnarray}
(U, V) &=& \frac {(\sinh(2u), \sinh(2v))} {\cosh(2u) + \cosh(2v)}, \label{2.29}
\\
u &=& \frac 1 4 \left(\ln \frac {1 + U + V} {1 - U - V} + \ln \frac {1 + U - V}
{1 - U + V}\right), \label{2.30} \\
v &=& \frac 1 4 \left(\ln \frac {1 + U + V} {1 - U - V} - \ln \frac {1 + U - V}
{1 - U + V}\right), \label{2.31}
\end{eqnarray}
and the metric becomes
\begin{equation}\label{2.32}
{\rm d} s^2 = \frac {f^2 \left({\rm d} V^2 - {\rm d} U^2\right)} {\left[1 - (U +
V)^2\right] \left[1 - (U - V)^2\right]} - r^2(U, V) \left({\rm d} \vartheta^2 +
\sin^2 \vartheta {\rm d} \varphi^2\right).
\end{equation}
The horizons have the equations $U = \pm V$, while the infinities of $p$ and $q$
become the four straight line segments $P = U + V = \pm 1$, $Q = U - V = \pm 1$.
Note from (\ref{2.29}) that $(u > 0) \Longleftrightarrow (U > 0)$, $(u = 0)
\Longleftrightarrow (U = 0)$, and the same is true for the pair $(v, V)$.

\begin{figure}[h]
 \begin{center}
 ${}$ \\[-9cm]
 \includegraphics[scale=0.8]{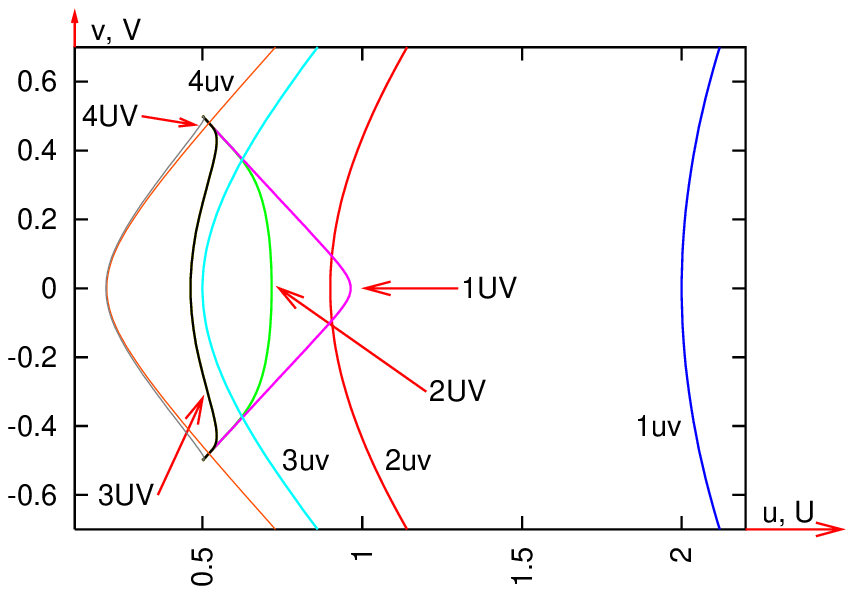}
 \caption{
 \label{rconst}
 \footnotesize
Comparison of the graphs of the functions ${\cal F} = \{u(t), v(t)\}$ and ${\cal
G} = \{U(t), V(t)\}$ given by (\ref{2.33}) -- (\ref{2.35}) for different
distances between the vertices of ${\cal F}$. See the explanations in the text.
Note that all $\{U(t), V(t)\}$ graphs run between $V = -0.5$ and $V = +0.5$.}
 \end{center}
 \end{figure}

Before we construct the diagram of the maximal analytic extension of the RN
spacetime (i.e. transform it piecewise to the $(U, V)$ coordinates) we have to
point out that the form of this diagram in earlier papers (e.g. \cite{Cart1973})
and textbooks (e.g. \cite{PlKr2024}) is not universally correct.\footnote{In
Refs. \cite{MTWh1973} and \cite{HaEl1973} the image of the singularity at $r =
0$ was drawn purely schematically, with no intention to show its real shape.}
The two functions ${\cal F} \df \{u(t), v(t)\}$ and ${\cal G} \df \{U(t),
V(t)\}$ at constant $r$ are given by the parametric equations
\begin{eqnarray}
u(t) &=& d \cosh(t), \qquad v(t) = d \sinh(t), \qquad d \df
{\rm e}^{\gamma r^*(r)}, \label{2.33} \\
U(t) &=& \sinh(2 u(t))/M(t), \qquad V(t) = \sinh(2 v(t))/M(t), \label{2.34} \\
M(t) &\df& \cosh(2 u(t)) + \cosh(2 v(t)).  \label{2.35}
\end{eqnarray}
Curve ${\cal F}$ is a hyperbola that degenerates to a pair of straight lines in
the limit of $d \to 0$. The shape of curve ${\cal G}$ depends on the value of
$d$, which is determined by $m$ and $e$ via (\ref{2.9}) and (\ref{2.3}). For
large $d$ ($d = 2$ on curves 1uv and 1UV in Fig. \ref{rconst}) ${\cal F}$ and
${\cal G}$ have curvatures of opposite signs and are far from each other. At an
intermediate $d$ ($d = 0.9$ on curves 2uv and 2UV, this is the case considered
further on in this paper), the two curves come closer together, but still have
curvatures of opposite signs. At a smaller $d$ ($d = 0.5$ on curves 3uv and
3UV), the curvature of ${\cal G}$ changes sign along it. At a sufficiently small
$d$ ($d = 0.2$ on curves 4uv and 4UV in Fig. \ref{rconst}) ${\cal G}$ nearly
coincides with a segment of ${\cal F}$ and has the same sign of curvature nearly
all along. This is the case for which the schematic figures in the literature so
far were drawn. For the beginning, we will follow the tradition and construct
the extension of the RN spacetime with a small $d$ in (\ref{2.33}).

\begin{figure}
 \begin{center}
 \hspace*{30mm}
  \includegraphics[scale=0.7]{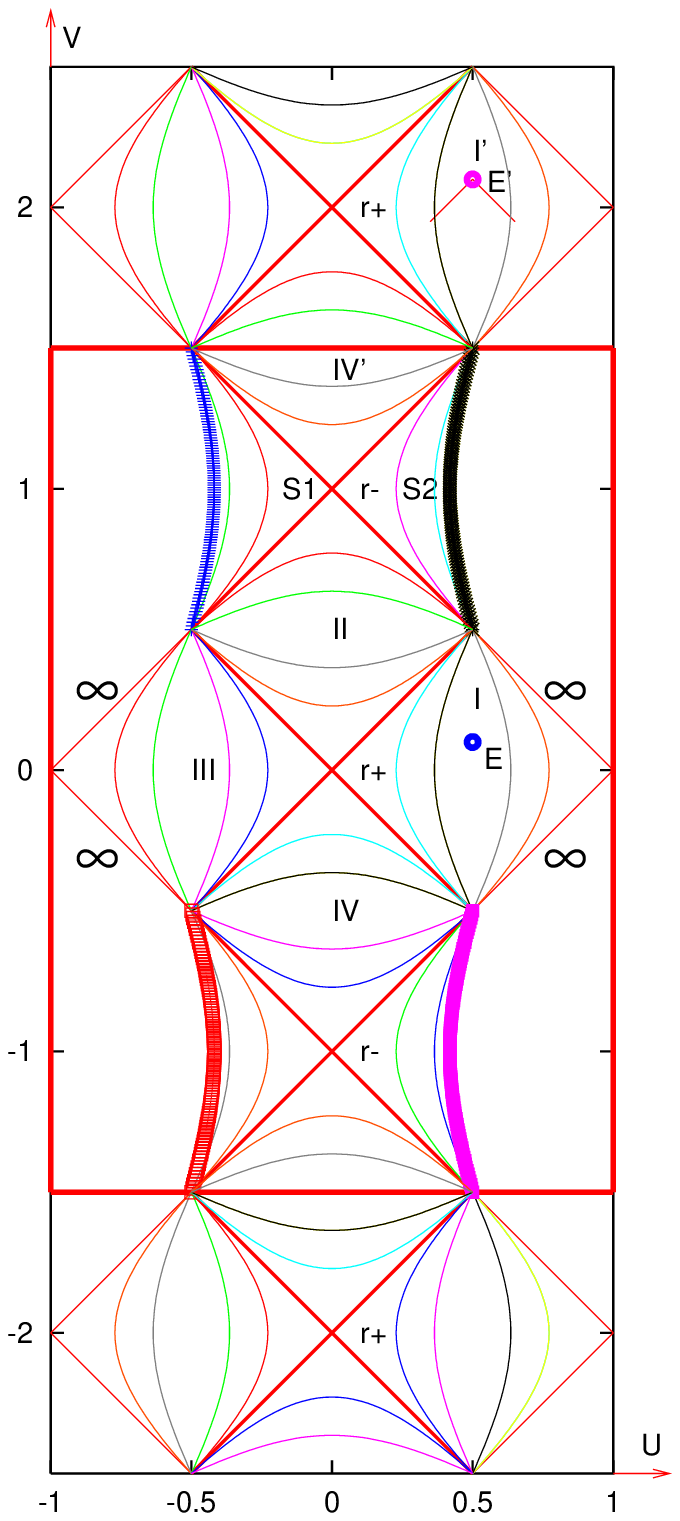}
 \caption{
 \label{reinormax}
 \footnotesize
The conformal diagram of the maximally extended Reissner--Nordstr\"{o}m
spacetime with $e^2 < m^2$ and with a small $d$ in (\ref{2.33}). The upper and
lower tunnel between the singularities can be identified, then sector I$'$ would
coincide with sector I, and event E$'$ in sector I$'$ (shown together with its
past light cone, PLC) would coincide with event E in sector I. Can a message be
sent from E to the PLC of E$'$? See the text for more explanation.}
 \end{center}
 \end{figure}

We can proceed from a point E in the $r > r_+$ region back in time along a $q =$
constant null geodesic and cross the spurious singularity $r = r_+$ at $p = 0$,
or to the future along a $p = $ constant null geodesic and cross $r = r_+$ at $q
= 0$. By extending these two kinds of null geodesics, we cover sectors I, II and
IV of Fig. \ref{reinormax}.\footnote{This figure first appeared in the papers by
Carter, see in particular Ref. \cite{Cart1973}.} By sending null geodesics back
in time from sector II and to the future from sector IV we cover sector III. In
the $r_- < r < r_+$ regions, we transform the $(u, v)$ coordinates continued
from sector I to such $(u', v')$ that cancel the $r = r_-$ spurious singularity.
The following remark will require attention in the subsequent sections:
\begin{eqnarray}\label{2.36}
&&{\rm We\ draw\ the\ conformal\ diagrams\ in\ } (U', V')\ {\rm corresponding\
to\ } (u', v') \nonumber \\
&&{\rm in\ such\ a\ way\ that\ their\ images\ of}\ r = 0\ {\rm coincide\
with\ the}\ r = 0\ {\rm sets\ } \nonumber \\
&& {\rm of\ the\ } (U, V)\ {\rm diagram\ and\ {\it this\ involves\ a\
transformation}. }
\end{eqnarray}
We will present this transformation in Sec. \ref{shaper}. The $r = r_-$ horizons
are again straight lines. By continuing the extensions and patching together
their results, we arrive at the manifold shown in Fig. \ref{reinormax}.

The thin straight segments in Fig. \ref{reinormax} are the images of the null
infinities, where $r \to \infty$. Their endpoints are the timelike and spacelike
infinities. The thin hyperbola segments represent the $r =$ constant lines; they
are timelike for $r > r_+$ and $r < r_-$ and spacelike for $r_- < r < r_+$. The
thicker straight segments are the horizons at $r = r_{\pm}$. The hatched
hyperbola segments represent the true singularities at $r = 0$. Radial null
geodesics (not shown) would be straight lines parallel to the spurious
singularities.

Re the meaning of `represent': The arcs in Fig. \ref{reinormax} are an
illustrative scheme, not an exact image. The shapes and positions in Fig.
\ref{reinormax} of the exact images of lines of constant $r$ depend on the
coordinates used and on the range of the $m$ and $e$ parameters. The schematic
presentation thus avoids complications, but see Sec. \ref{shaper} and Figs.
\ref{drawgeo}, \ref{drawmany} and \ref{drawback}.

The upper tunnel between the true singularities within the thick rectangle is a
copy of the lower one. We can identify the two tunnels, thus making the extended
manifold cyclic in time. Alternatively, we can continue to send null geodesics
to the future from the upper tunnel and to the past from the lower tunnel, to
obtain an infinite chain of asymptotically flat regions and tunnels. A natural
suspicion is that the identification may result in an acausal spacetime, in
which one could send signals (by means of geodesics) to the future and receive
them from the past before they were sent. This would lead to paradoxes such as
sending the order ``do not emit any message if you receive this one'', and yet
receiving it before it was sent. In the present paper it is shown that the
acausality does or does not occur depending on the position of the origin of the
geodesic.

\section{The geodesic equations in the $(u, v)$ coordinates}\label{geode}

\setcounter{equation}{0}

The geodesic equations for the RN metric in the coordinates of (\ref{2.1}) --
(\ref{2.2}) have been presented and partly integrated in previous
publications,\footnote{Actually, some of those publications discussed the orbits
of charged particles, which are not geodesics. But they become geodesic in the
limit of zero charge of the particle.} see e.g. Ref. \cite{PlKr2024} (Sec.
14.16). For each single timelike or null geodesic, the $(\vartheta, \varphi)$
coordinates can be adapted so that the whole geodesic lies in the $\vartheta =
\pi / 2$ hypersurface, and then $\varphi$ along it obeys
\begin{equation}\label{3.1}
\dril {\varphi} s = J_0 / r^2,
\end{equation}
where $J_0$ is an arbitrary constant and $s$ is an affine parameter. The
geodesic is radial when $J_0 = 0$. Obtaining these results takes up 2 of the 4
geodesic equations, and the remaining 2 have the following first integrals:
\begin{eqnarray}
\phi\ \dril t s &=& \Gamma = {\rm constant}, \label{3.2} \\
\dril r s &=& \sigma \sqrt{\Gamma^2 - E \phi}, \label{3.3} \\
E &\df& \varepsilon + {J_0}^2 / r^2, \label{3.4}
\end{eqnarray}
where $\Gamma > 0$ ($< 0$) on future- (past-) directed geodesics in sector
I,\footnote{In sectors where $\phi < 0$, Eq. (\ref{3.2}) shows that $t$
decreases along the geodesic when $\Gamma > 0$, but it is a space coordinate
there.} $\sigma = +1$ ($-1$) on outgoing (ingoing) geodesics, $\varepsilon =
{\rm constant} > 0$ on timelike geodesics and $\varepsilon = 0$ on null
geodesics. With $\varepsilon = 1$ and $\Gamma^2 > 1$, the orbit determined by
(\ref{3.2}) -- (\ref{3.4}) is hyperbolic ($r$ can go to infinity, and $\lim_{r
\to \infty} |\dril r s| > 0$), with $\Gamma^2 = 1 = \varepsilon$ it is parabolic
($r \to \infty$ allowed, but $\lim_{r \to \infty} \dril r s = 0$), with
$\Gamma^2 < 1 = \varepsilon$ it is elliptic ($r < \infty$ permanently).

Equations (\ref{3.2}) -- (\ref{3.4}) imply the general first integral of
geodesic equations:
\begin{equation}\label{3.5}
\phi \left(\dr t s\right)^2 - \frac 1 {\phi}\ \left(\dr r s\right)^2 - \frac
{{J_0}^2} {r^2} = \varepsilon.
\end{equation}

Taking (\ref{2.14}) for $t$ and $r$ we find that in the $(u, v)$ coordinates
(\ref{3.2}) -- (\ref{3.3}) become
\begin{eqnarray}
&& \frac {\phi} {\gamma \left(v^2 - u^2\right)} \left(v \dr u s - u \dr v
s\right) = \Gamma, \label{3.6} \\
&& \frac {\phi} {\gamma \left(v^2 - u^2\right)} \left(v \dr v s - u \dr u
s\right) = \sigma \sqrt{\Gamma^2 - E \phi}. \label{3.7}
\end{eqnarray}
The same formulae (\ref{3.6}) -- (\ref{3.7}) follow when we use (\ref{2.17})
instead of (\ref{2.14}) for calculating $\dril t s$ and $\dril r s$.
Consequently, when we extend a geodesic from sector I in Fig. \ref{reinormax}
into sector II, we can continue to use Eqs. (\ref{3.6}) -- (\ref{3.7}) and
conclusions from them.

Still the same formulae (\ref{3.6}) -- (\ref{3.7}) result in sector S1.

Equations (\ref{3.6}) -- (\ref{3.7}) are equivalent to
\begin{eqnarray}
\dr u s &=& \frac {\gamma} {\phi}\ \left[\Gamma v + \sigma u\ \sqrt{\Gamma^2 - E
\phi}\right], \label{3.8} \\
\dr v s &=& \frac {\gamma} {\phi}\ \left[\Gamma u + \sigma v\ \sqrt{\Gamma^2 - E
\phi}\right].\label{3.9}
\end{eqnarray}
Since $\phi < 1$ for $r_- > e^2 / (2m) < r < \infty$, it follows that $\Gamma^2
- E \phi \geq 0$ when $0 < E \leq \Gamma^2$ throughout sectors I and II and in a
part of sector S1. The locus of $\Gamma^2 - E \phi = 0$ is the TP for $r(s)$ as
seen from (\ref{3.3}). For radial ($J_0 = 0$) timelike ($\varepsilon = +1$)
geodesics the value of $r$ at the TP is
\begin{equation}\label{3.10}
\textcolor[rgb]{0.00,0.00,1.00}{r_{{\rm tp} \pm} = \frac 1 {\Gamma^2 - 1}\
\left(- m \pm \sqrt{m^2 - e^2 + \Gamma^2 e^2}\right).}
\end{equation}
\setcounter{footnote}{6} The sign is $+$ for $\Gamma^2 > 1$, and both signs are
allowed when $\Gamma^2 < 1$.\footnote{\textcolor[rgb]{0.98,0.00,0.00}{For
$\Gamma^2 < 1$ we have $r_{{\rm tp}+} < r_-$, $r_{{\rm tp}-} > r_+$ and $r_{{\rm
tp}+} \llim{\Gamma^2 \to 1} e^2/(2m)$.}}

For numerical calculation, it will be  convenient to take $v$ as the independent
variable and calculate $u(v)$. Then, from (\ref{3.8}) -- (\ref{3.9})
\begin{equation}\label{3.11}
\dr u v = \frac {\Gamma v + \sigma u \sqrt{\Gamma^2 - E \phi}} {\Gamma u +
\sigma v \sqrt{\Gamma^2 - E \phi}}.
\end{equation}
For radial null ($E = 0$) geodesics, (\ref{3.11}) implies $\dril u v = \pm 1$,
i.e.
\begin{equation}\label{3.12}
v \pm u = D = {\rm constant}.
\end{equation}
On outgoing radial null geodesics, from (\ref{3.12}) and (\ref{2.29}) we have
\begin{equation}\label{3.13}
V - U = \frac {\sinh(2v) - \sinh(2u)} {\cosh(2u) + \cosh(2v)} \equiv \frac
{\sinh(v - u)} {\cosh(v - u)} = \tanh D.
\end{equation}

For geodesics with $\sigma = -1$ that cross the horizons at $r = r_{\pm}$, where
$u = \pm v$ and $\phi = 0$, Eq. (\ref{3.11}) becomes 0/0. It is shown in
Appendix \ref{numer} how to deal with this problem.

Note that (\ref{3.11}) is invariant under the transformation $(v, u) = (u',
v')$.

While integrating (\ref{3.11}) numerically, we will need the value of $r$ at
each step to calculate $\phi(r)$. Given $u$, $v$ and $\gamma$, (\ref{2.14})
determines $r^*$. The numerical calculation of the corresponding $r$ from
(\ref{2.9}) poses a problem that is explained and solved in Appendix
\ref{bisec}.

\section{Transformations between different coordinates}\label{thepro}

\setcounter{equation}{0}

On lines of constant $r$, as seen from (\ref{2.14}) -- (\ref{2.19}), $u^2 - v^2
= C =$ constant. The sign of $C$ depends on the region of the $(u, v)$ surface.
The following observation will be useful in further calculations:

{\bf Lemma 4.1}

Equation (\ref{2.29}) implies that $\dril V v > 0$ along a line of constant $r$
when $C > 0$.

For the proof see Appendix \ref{dVbydv}.

The conclusion is that after the transformation $(u, v) \to (U, V)$ given by
(\ref{2.29}) the ordering of events by the value of $V$ along a line of constant
$r$ agrees with the ordering by the value of $v$ in those sectors of Fig.
\ref{reinormax} in which $u^2 - v^2 > 0$. In the other sectors the two orderings
are not necessarily consistent with each other.

By the same method follows

{\bf Lemma 4.2}

Equation (\ref{2.29}) implies that $\dril U u > 0$ along a line of constant $r$
in regions where $u^2 - v^2 < 0$. $\square$

The ordering by the values of $v$ agrees with the ordering by the values of $t$
only when $\gamma > 0$ in (\ref{2.13}) and (\ref{2.19}). With $\gamma < 0$, the
two orderings are opposite. With (\ref{2.16}) and (\ref{2.18}), the relation
between the two orderings is still more complicated.

{}From now on we will denote the $(U, V)$ coordinates in Fig. \ref{reinormax} by
$({\cal U}, {\cal V})$. They do not coincide with the internal $(U, V)$
coordinates of most sectors. In the rectangle consisting of sectors II and S1
the $({\cal U}, {\cal V})$ are shifted by $(\delta {\cal U}, \delta {\cal V}) =
(-0.5, +0.5)$ with respect to the internal $(U, V)$. The shifts in the other
places can now be easily deduced.

While calculating the geodesics we will have to transform between the $\gamma_1$
coordinates $(u, v)$ and the $\gamma_2$ coordinates $(u', v')$ somewhere in
sectors II and IV$'$ in Fig. \ref{reinormax}. In order to derive the
transformation formulae we have to recall that square II in the figure is in
truth sector II for the $\gamma_1$ coordinates (where $u^2 - v^2 < 0$) overlaid
on sector I for the $\gamma_2$ coordinates (where $u'^2 - v'^2 > 0$). The
transformation formulae follow from the condition that the pairs $(u, v)$ and
$(u', v')$ correspond to the same pair $(t, r)$, so
\begin{eqnarray}
{\rm e}^{2r^*} &=& \left(\frac {v^2 - u^2} {4 A^2}\right)^{1 / \gamma_1} =
\left(\frac {u'^2 - v'^2} {4 A^2}\right)^{1 / \gamma_2}, \label{4.1} \\
t &=& \frac 1 {\gamma_1} {\rm artanh} (u/v) = \frac 1 {\gamma_2} {\rm artanh}
(v'/u'), \label{4.2}
\end{eqnarray}
where $A > 0$. Using the identity ${\rm artanh}\ x \equiv \frac 1 2\ \ln \frac
{1 + x} {1 - x}$ and denoting
\begin{equation}\label{4.3}
a \df \left(r_- / r_+\right)^2 \equiv - \gamma_1 / \gamma_2
\end{equation}
Eqs. (\ref{4.1}) -- (\ref{4.2}) lead to\footnote{In fact, (\ref{4.1}) --
(\ref{4.2}) determine $(u + v)^2$ and $(u - v)^2$. But there is no sign
ambiguity in calculating $(u + v)$ and $(u - v)$ therefrom because, with the
assumed $A > 0$ and ${\rm e}^{\gamma r^*} \geq 0$, Eqs. (\ref{2.13}) guarantee
that $u \pm v \geq 0$ in sector I (so $u' \pm v' \geq 0$ in (\ref{4.4}) --
(\ref{4.5})) while (\ref{2.16}) guarantee that $v \pm u \geq 0$ in sector II.
The same applies to calculating $(u' \pm v')$ for (\ref{4.6}) -- (\ref{4.7}).}
\begin{eqnarray}
u &=& \frac 1 2\ (2A)^{1 + a} \left[(u' + v')^{-a} - (u' - v')^{-a}\right],
\label{4.4} \\
v &=& \frac 1 2\ (2A)^{1 + a} \left[(u' + v')^{-a} + (u' - v')^{-a}\right].
\label{4.5}
\end{eqnarray}
The inverse formulae are
\begin{eqnarray}
u' &=& \frac 1 2\ (2A)^{1 + 1/a} \left[(v + u)^{-1/a} + (v - u)^{-1/a}\right],
\label{4.6} \\
v' &=& \frac 1 2\ (2A)^{1 + 1/a} \left[(v + u)^{-1/a} - (v - u)^{-1/a}\right].
\label{4.7}
\end{eqnarray}
When $v + u \to 0$ (i.e. the point of coordinates $(u, v)$ approaches the
horizon $r = r_+$), both $u'$ and $v'$ go to $+ \infty$, i.e. the point $(u',
v')$ approaches $r = r_-$ in sector II.

Now we can calculate $\dril {u'} {v'}$ (see Appendix \ref{dubydv}). The result
\begin{equation}\label{4.8}
\dr {u'} {v'} = \frac {\Gamma v' + \sigma u' \sqrt{\Gamma^2 - E \phi}} {\Gamma
u' + \sigma v' \sqrt{\Gamma^2 - E \phi}}
\end{equation}
is a copy of (\ref{3.11}). This is consistent with the fact that on proceeding
from sector I to sector II $u$ and $v$ interchange, while (\ref{3.11}) is
invariant under such a transformation.

For radial null geodesics on which $v - u = D$, Eqs. (\ref{4.6}) -- (\ref{4.7})
give
\begin{equation}\label{4.9}
v' - u' = - (2A)^{1 + 1/a} D^{-1/a},
\end{equation}
so $v' - u'$ is also constant, but different from $D$.

We also need the transformations analogous to (\ref{4.4}) -- (\ref{4.7}) for
past-directed radial geodesics proceeding from sector I$'$ to sector IV$'$. This
is equivalent to doing the same transformation for geodesics going from sector I
to sector IV. A comparison of (\ref{2.16}) with (\ref{2.18}) shows that in
sector IV we have $v \leq 0$ and $|u| \leq |v|$, so $v + u \leq 0$ and $v - u
\leq 0$. Therefore, $\sqrt{(v + u)^2} = - v - u$ and $\sqrt{(v - u)^2} = u - v$.
In consequence of this
\begin{eqnarray}\label{4.10}
&& {\rm The\ desired\ formulae\ follow\ from\ (\ref{4.6})\ -\ (\ref{4.7})}
\nonumber \\
&& {\rm by\ replacing}\  v + u \to - v - u, \quad v - u \to u - v.
\end{eqnarray}

\section{Lines of constant $r$ in Fig. \ref{reinormax}}\label{shaper}

\setcounter{equation}{0}

Sector I of the $\gamma_2$ coordinates, as already mentioned, coincides in Fig.
\ref{reinormax} with sector II of the $\gamma_1$ coordinates. In the former,
$u^2 - v^2 > 0$ and the arcs of constant $r$ are standing vertically as in
sector I. In the latter, $u^2 - v^2 < 0$ and these arcs are lying horizontally
as in sector II of the figure.

The lines of constant $r = r_0$ are determined as follows: given $r_0$ and
$\gamma$, we calculate $r^*(r_0)$ from (\ref{2.9}), then $d(r_0) \df 2A {\rm
e}^{\gamma r^*(r_0)}$, and we use (\ref{2.33}) -- (\ref{2.35}). The $({\cal U},
{\cal V})$ coordinates of this line in the figure are calculated by applying the
relevant shifts. The shape and position of the resulting line depends on the
value of $d$, see Fig. \ref{rconst}. In the $(u, v)$ coordinates with $u^2 - v^2
< 0$, Eq. (\ref{2.33}) applies with $u$ and $v$ interchanged.

\begin{figure}
 \begin{center}
 ${}$ \\[-1cm]
 \includegraphics[scale=0.7]{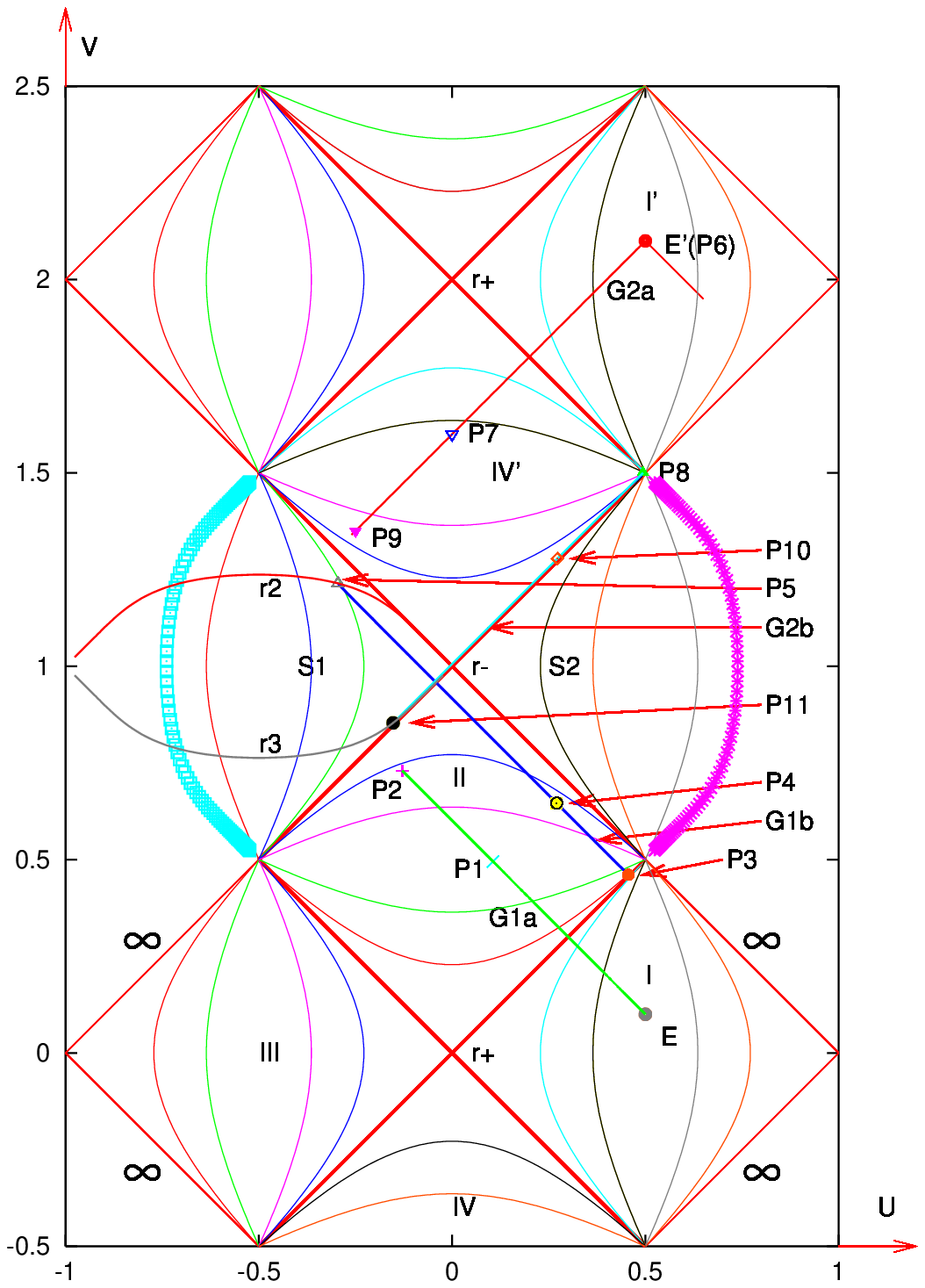}
 \caption{
 \label{drawgeo}
 \footnotesize
A radial timelike geodesic running from event E in sector I to the turning point
$P_5$ on the arc $r_2$. The hatched arcs are the exact images of the singularity
at $r = 0$. The meaning of the other elements in this figure is gradually
explained as the text proceeds. }
 \end{center}
 \end{figure}

Our further reasoning will be carried out within the subset of Fig.
\ref{reinormax} shown in Fig. \ref{drawgeo}. In the latter, the arcs of the
singularity at $r = 0$ are adapted to the numerical values of $r_-$ and $r_+$
that are introduced in Sec. \ref{numcal} -- hence the change explained in Fig.
\ref{rconst} and in the associated segment of the text. The lines of constant
$r$ are drawn in sectors I and S1 of Fig. \ref{drawgeo} as if they had $u^2 -
v^2 > 0$. However, when we follow a geodesic that begins in sector II in the
$\gamma_2$ coordinates, it arrives in sector S1 with $u^2 - v^2 < 0$ (the sign
of $u^2 - v^2$ changes when crossing $r = r_-$) and its arc of $r = r_{\rm tp}$
(i.e. the locus of the TPs given by (\ref{3.10})) lies horizontally -- it is
marked $r_2$ in Fig. \ref{drawgeo}. In the $(u, v)$ coordinates with $u^2 - v^2
< 0$ and $\gamma = \gamma_2 < 0$, Eq. (\ref{2.16}) shows that $u$ decreases when
$t$ increases. So, on the $r_2$ arc, the time-ordering of events is right to
left, i.e. the events with larger ${\cal U}$ (the rightmost ones) are earlier
than those with smaller ${\cal U}$ (recall Lemma 4.2).

There are more elements in Fig. \ref{drawgeo} than can be explained at this
stage; they will be explained further on as we go.

\begin{figure}[h]
 \begin{center}
 ${}$ \\[-4.5cm]
 \includegraphics[scale=0.8]{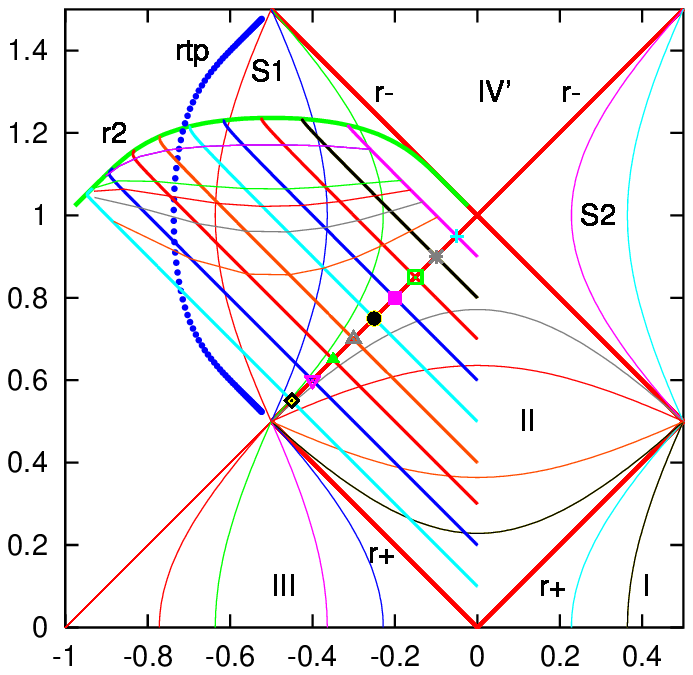}
 \caption{
 \label{drawmany}
 \footnotesize
The future-directed radial timelike geodesics that go off in sector II in the
$\gamma_2$ coordinates with $u^2 - v^2 < 0$ reach their turning points at the
arc $r_2$. The locus of the singularity at $r = 0$ seems to coincide with $r_2$
at the scale of this figure, but lies above the latter. The $r$ along the
geodesics goes through the value $r_-$ at the marked points. The dotted arc
marked rtp is the locus of turning points in sector S1 in the $\gamma_2$
coordinates with $u^2 - v^2 > 0$, see Fig. \ref{drawback}. The thin lines
transversal to the goedesics are loci of constant $r$, from bottom to top they
are $r = 0.89, 0.88, 0.87, 0.86$ and $r = 0.8$.}
 \end{center}
 \end{figure}

Figure \ref{drawmany} shows a collection of future-directed radial timelike
geodesics that go off in sector II in the $\gamma_2$ coordinates with $u^2 - v^2
< 0$. Their initial points all have ${\cal U} = 0$, while their ${\cal V} = 0.1,
0.2, \dots, 0.9$. They cross the $r = r_-$ horizon at the line $r_-$ in the
figure. The locus of their TPs lies at $r_2$ because the $(u, v)$ coordinates
used here are not the internal $(u, v)$ of sector S1. The arc of $r = 0$ in
these coordinates lies above $r_2$, but at the scale of Fig. \ref{drawmany} is
indistinguishable from $r_2$.

\begin{figure}[h]
 \begin{center}
 ${}$ \\[-5cm]
 \includegraphics[scale=0.7]{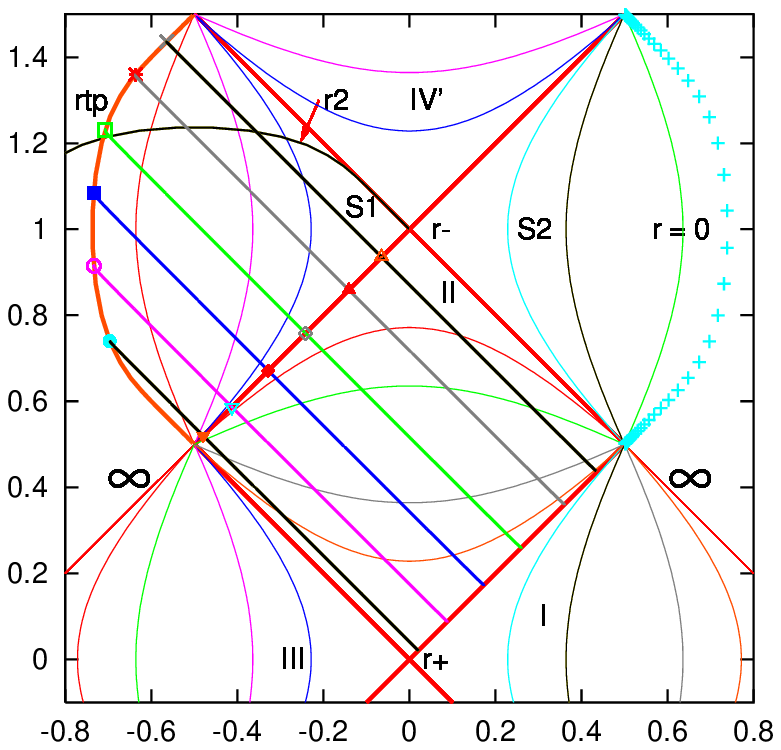}
 \caption{
 \label{drawback}
 \footnotesize
The past-directed radial timelike geodesics that go off the arc rtp -- the locus
of their turning points in the $\gamma_2$ coordinates with $u^2 - v^2 > 0$. See
the text for explanations.}
 \end{center}
 \end{figure}

Figure \ref{drawback} shows a collection of past-directed radial timelike
geodesics that go off the arc $r = r_{\rm tp}$ in sector S1 in the $\gamma_2$
coordinates with $u^2 - v^2 > 0$.\footnote{This is a different collection from
that in Fig. \ref{drawmany}. Their initial points are determined from
(\ref{2.33}) -- (\ref{2.35}) with the value of $d$ that follows from (\ref{6.1})
-- (\ref{6.2}) and $\gamma = \gamma_2$; they have, from top to bottom, $t = 2.0,
1.2, 0.6, 0.2, -0.2, -0.7$.} They cross the horizon $r = r_-$ at the marked
points. In the figure, they come close to $r = r_+$ but cannot reach it because
in the $\gamma_2$ coordinates the set $r = r_+$ lies at infinity -- see
(\ref{2.25}) and (\ref{2.15}). Figures \ref{drawmany} and \ref{drawback}
illustrate the remark (\ref{2.36}): to fit in Fig. \ref{drawgeo}, the image in
Fig. \ref{drawmany} has to be transformed into that in Fig. \ref{drawback}; see
below for the transformation. The left arc of $r = 0$ in these coordinates lies
to the left of $r = r_{\rm tp}$, but at the scale of Fig. \ref{drawback} the two
are indistinguishable; the exact second image of $r = 0$ is shown on the right.

\begin{figure}[h]
 \begin{center}
 ${}$ \\[-4cm]
 \includegraphics[scale=0.7]{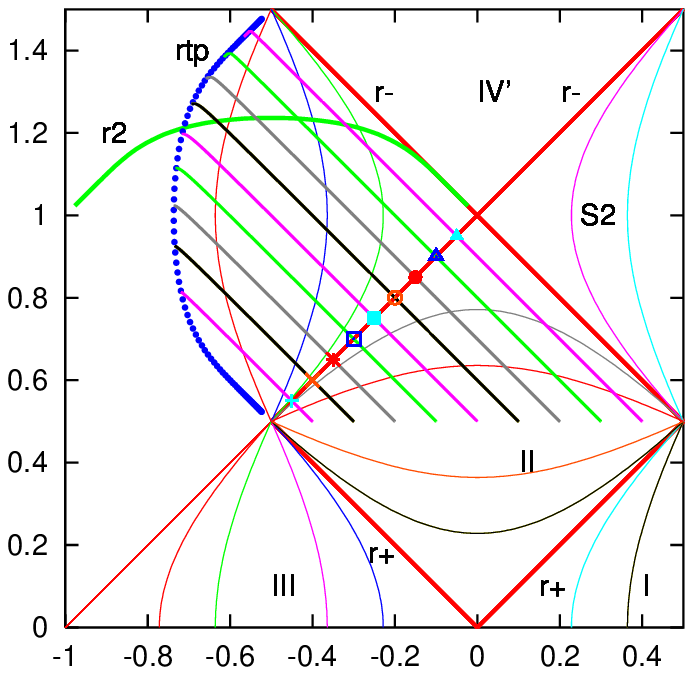}
 \caption{
 \label{drawmanytr}
 \footnotesize
The transformation of Fig. \ref{drawmany} by (\ref{5.1}) with the shift
(\ref{5.2}) applied to the result. This graph is consistent with Fig.
\ref{drawback}.}
 \end{center}
 \end{figure}

The configuration in sectors II and S1 of Fig. \ref{drawback} is a
mirror-reflection in the line $V = - U + 0.5$ of that in Fig. \ref{drawmany},
where the $(U, V)$ are the internal coordinates of these
sectors.\footnote{Except that the two sets of geodesics are different. See Fig.
\ref{drawmanytr} for an exact comparison.} A reflection of a point of
coordinates $(x, y)$ in the line $y = - x$ would be described by the equations
$(x', y') = (-y, -x)$. Consequently, the reflection in the line $y = - x + b$ is
described by the equations $(x', y') = (-y + b, -x + b)$, so
\begin{equation}\label{5.1}
(U', V') = (-V + b, -U + b).
\end{equation}
To place the result of this reflection in the frame of Fig. \ref{drawmany} we
have to shift it by
\begin{equation}\label{5.2}
(\delta {\cal U}, \delta {\cal V}) = (-0.5, 0.5).
\end{equation}
The transformation of Fig. \ref{drawmany} by (\ref{5.1}) -- (\ref{5.2}) is shown
in Fig. \ref{drawmanytr} -- it is consistent with Fig. \ref{drawback}. This is
the transformation implied in (\ref{2.36}).

The transformation in the $(u, v)$ coordinate surface corresponding to
(\ref{5.1}) is
\begin{equation}\label{5.3}
\left(\begin{array}{l} u' \\ v' \end{array}\right) = \frac 1 4\ \ln \left[\frac
{{\rm e}^{- u - v} + 2a \cosh (u + v)} {{\rm e}^{u + v} - 2a \cosh (u +
v)}\right] \pm \frac 1 2\ (u - v).
\end{equation}

\section{A timelike future-directed geodesic that does not break
causality}\label{numcal}

\setcounter{equation}{0}

The problem we now wish to solve is the following. Let E be an event in sector I
of Fig. \ref{drawgeo}. Let E$'$ be the copy of E in sector I$'$ (which is the
first future copy of sector I). Is it possible to send from E a future-directed
timelike or null geodesic $G$ such that it goes through the tunnel between the
singularities and enters the past light cone of E$'$? If this is possible, then
the identification described in Sec. \ref{RNintro} indeed leads to causality
violation: E can send a message to its causal past.

$G$ cannot be radial null: in Fig. \ref{drawgeo} it would consist of straight
segments running parallel to the event horizons and would hit the singularity at
left (see Eq. (\ref{3.3}) -- there is no TP when $E = 0$). To enter sector I$'$,
it would have to be reflected somewhere within the tunnel and the RN geometry
does not contain any such mirror.

Can $G$ be radial timelike? We will deal with this question now. An example will
be presented of a radial timelike geodesic proceeding from event E in sector I
of Fig. \ref{drawgeo} to the locus of the TPs. For the parameters of the RN
metric and of the geodesic we choose:
\begin{eqnarray}
r_+ &=& 1.0, \qquad r_- = 0.9, \qquad 2A = 1, \label{6.1} \\
\Gamma &=& 1.1, \qquad E = 1.0. \label{6.2}
\end{eqnarray}

The computation (at double precision in Fortran 90) proceeds as follows:

1. We choose the initial point E in sector I. Its $({\cal U}, {\cal V})$
coordinates in Fig. \ref{drawgeo} are
\begin{equation}\label{6.3}
({\cal U}_0, {\cal V}_0) = (0.5, \ \ 0.1),
\end{equation}
and its internal $(U_0, V_0)$ in sector I are the same. The corresponding $(u,
v, r)$ are\footnote{$(u_0, v_0)$ are calculated using (\ref{2.30}) --
(\ref{2.31}), $r_0$ is calculated by solving the first of (\ref{2.14}) for $r$.}
\begin{eqnarray}\label{6.4}
(u_0, v_0) &=& (0.55839805537677356, \quad 0.13474912518317173), \nonumber \\
r_0 &=& 1.0598002056721780;
\end{eqnarray}
the high precision will ensure that the calculated effect is larger than a
numerical error.

2. We integrate (\ref{3.11}) off the point E. The integration proceeds towards
decreasing $u$ when $\sigma = -1$ (see (\ref{3.8}) -- (\ref{3.9})), and towards
the future when $\Gamma > 0$ (see (\ref{3.2})). We thus launch the
future-directed ingoing radial timelike geodesic G1a. We continue the
integration up to point $P_2$, at which $r = 0.94$, whose coordinates in Fig.
\ref{drawgeo} are
\begin{equation}\label{6.5}
({\cal U}_2, {\cal V}_2) = (-0.12857138214262273, \quad 0.72967049671654927).
\end{equation}
G1a crosses $r = r_+$ smoothly using (\ref{3.11}). As (\ref{2.9}) shows,
$\lim_{r \to r_-} r^* = + \infty$, $\lim_{r \to r_+} r^* $ $= - \infty$ and
$\dril {r^*} r = 1 / \phi < 0$ for $r_- < r < r_+$, so with $\gamma = \gamma_1$
the function $r^*(r)$ is monotonic in this range and $r$ at $P_2$ is uniquely
determined.

 \begin{figure}[h]
 \begin{center}
 ${}$ \\[-2.5cm]
 \includegraphics[scale=0.7]{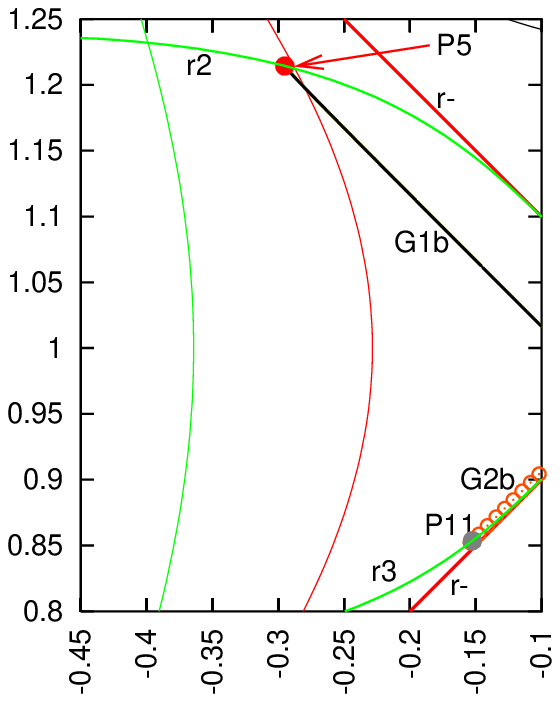}
 \caption{
 \label{lup5}
 \footnotesize
Closeup view on the neighbourhood of points $P_5$ (lying on $r_2$) and $P_{11}$
(lying on $r_3$) in Fig. \ref{drawgeo}. See the text for more explanation.}
 \end{center}
 \end{figure}

3. At $r = 0.97$, which occurs at point $P_1$, we transform its $(u, v)$ by
(\ref{4.6}) -- (\ref{4.7}) to $(u', v')$ -- the coordinates of $P_3$ (the image
of $P_1$). The second segment of $G_1$, denoted G1b, takes off at $P_3$ whose
coordinates in the figure are
\begin{equation}\label{6.6}
({\cal U}_3, {\cal V}_3) = (0.45594056360076862, \quad 0.46086405038409167).
\end{equation}
These differ by $(\delta {\cal U}, \delta {\cal V}) = (-0.5, +0.5)$ from the
internal $(U, V)$ of sector I of the $\gamma_2$ coordinates.

4. We integrate (\ref{4.8}) off $P_3$ with $\sigma = -1$. Between $r(P_1)$ and
$r(P_2)$, G1a and G1b run side by side.

5. A radial timelike geodesic cannot hit the singularity at $r = 0$. If it
enters the $r < r_-$ region, then it must have a TP at the $r_{\rm tp} < r_-$
given by (\ref{3.10}) (see Ref. \cite{PlKr2024}, Sec. 14.16). This is point
$P_5$ in sector S1, whose coordinates in the figure are
\begin{equation}\label{6.7}
({\cal U}_5, {\cal V}_5) = (-0.29505173312860367, \quad 1.2143536937009773).
\end{equation}
Figure \ref{lup5} shows a closeup view on the neighbourhood of $P_5$.

\section{The past light cone of E$'$}\label{PLCE}

\setcounter{equation}{0}

We continue the enumeration of the phases of calculation from Sec. \ref{numcal}.

6. Now we move to event E$'$ (also denoted $P_6$), whose coordinates in Fig.
\ref{drawgeo} are
\begin{equation}\label{7.1}
({\cal U}_6, {\cal V}_6) = (0.5, \ \  2.1),
\end{equation}
but its internal $(U, V)$ and $(u, v)$ are the same as those of E, given by
(\ref{6.3}) and (\ref{6.4}). We send from E$'$ the past-directed ingoing radial
{\it null} geodesic G2a, which is the radial generator of the past light cone of
E$'$. Its image in the figure can be calculated exactly (i.e. without numerical
computation) using (\ref{3.13}) and (\ref{6.3}), it obeys
\begin{equation}\label{7.2}
{\cal V} - {\cal U} = {\cal V}_0 - {\cal U}_0 = -0.4,
\end{equation}
and we follow it down to point $P_7$ in sector IV$'$ with coordinates in the
figure
\begin{equation}\label{7.3}
({\cal U}_7, {\cal V}_7) = (0.0, 1.6).
\end{equation}
The corresponding internal $(U_7, V_7)$ of $P_7$ in sector IV$'$ are $(0.0,
-0.4)$.

7. We transform the $(u_7, v_7)$ corresponding to (\ref{7.3}) to $(u'_7, v'_7)$
by (\ref{4.10}), and calculate the $(U'_7, V'_7) \df (U_8, V_8)$ coordinates of
point $P_8$ using (\ref{2.29}). We are now in sector IV$'$, so the $({\cal U},
{\cal V})$ coordinates in the figure are shifted with respect to the internal
$(U, V)$ of sector IV$'$ by $(\delta U, \delta V) = (-0.5, +1.5)$. The
coordinates of $P_8$ in the figure are\footnote{The value of ${\cal V}_8$
calculated by the plotting program gnuplot is exactly 1.5.}
\begin{equation}\label{7.4}
({\cal U}_8, {\cal V}_8) = (0.49380788638802, \quad 1.5).
\end{equation}
At the scale of Fig. \ref{drawgeo}, this point lies so near to the intersection
of the singularity arc on the right with $r = r_{\pm}$ that the two points are
indistinguishable. The internal $(U, V)$ coordinates of $P_8$ in sector IV$'$
are
\begin{equation}\label{7.5}
(U_8, V_8) = ({\cal U}_8 + 0.5, {\cal V}_8 - 1.5) = (0.99380788638802,\quad
0.0).
\end{equation}

8. From $P_8$ we issue the second segment of $G_2$, denoted G2b, that obeys the
equation $V - U = V_8 - U_8$ and continue it to the point $P_{11}$ of
intersection with the $r = r_{\rm tp}$ arc. At the scale of Fig. \ref{drawgeo},
the G2b segment looks to coincide with $r = r_-$. Just like G1b, the segment G2b
arrives in sector S1 with unadapted coordinates, so the arc of $r = r_{\rm tp}$,
denoted $r_3$, lies horizontally and is the mirror-reflection of $r_2$ in the
line ${\cal {\cal U}} = 1$.

The $({\cal U}, {\cal V})$ coordinates of $P_{11}$ in Fig. \ref{drawgeo} can be
calculated as follows. Let $V_8 - U_8 \df D = $ constant. Then, from
(\ref{3.13}), the equation of G2b in the $(u, v)$ coordinates is
\begin{equation}\label{7.6}
v - u = {\rm artanh}\ D \equiv \frac 1 2\ \ln \left(\frac {1 + D} {1 -
D}\right).
\end{equation}
{}From (\ref{2.33}) and the paragraph containing it adapted to $u^2 - v^2 < 0$
and $\gamma = \gamma_2$, the equation of the set $r = r_{\rm tp}$ is
\begin{equation}\label{7.7}
v^2 - u^2 = d^2(r_{\rm tp}) = 4 A^2 {\rm e}^{2 \gamma_2 r^*(r_{\rm tp})} \df
{d_0}^2.
\end{equation}
The solution of (\ref{7.6}) -- (\ref{7.7}) are the $(u, v)$ coordinates of
$P_{11}$,
\begin{equation}\label{7.8}
u = \frac 1 2\ \left({d_0}^2 / {\rm artanh}\ D - {\rm artanh}\ D\right) \qquad
v= \frac 1 2\ \left({d_0}^2 / {\rm artanh}\ D + {\rm artanh}\ D\right).
\end{equation}
The coordinates of $P_{11}$ in the figure are calculated from (\ref{2.29}) (with
the shifts applied):
\begin{equation}\label{7.9}
({\cal U}_{11}, {\cal V}_{11}) = (-0.15257743, \quad 0.85361468).
\end{equation}
The neighbourhood of $P_{11}$ is seen in Fig. \ref{lup5}. It is now clear that
$P_5$ lies to the future of the past light cone of E$'$ -- because its ${\cal
U}$ coordinate is smaller than that of $P_{11}$ (recall Lemma 4.2).
Consequently, no future-directed timelike or null geodesic that originates at
$P_5$ can enter the past light cone of E$'$. This concludes the proof for a
radial timelike geodesic emitted at E. $\square$

9. As a check of precision we extend the G2a segment down beyond $P_7$, to point
$P_9$ whose $({\cal U}, {\cal V})$ in the figure are
\begin{equation}\label{7.10}
({\cal U}_9, {\cal V}_9) = (-0.25, \ \ 1.35).
\end{equation}
We apply the calculation described in point 7 to the internal $(U, V)$ of sector
I$'$ corresponding to (\ref{7.10}), they are $(-0.25, -0.65)$. In this way we
find the coordinates of $P_{10}$ -- the image of $P_9$ under (\ref{4.10}). The
coordinates of $P_{10}$ in the figure are
\begin{equation}\label{7.11}
({\cal U}_{10}, {\cal V}_{10}) = (0.27258465, \quad 1.27877676).
\end{equation}

\section{A nonradial timelike geodesic}\label{norad}

\setcounter{equation}{0}

Let $G_J$ be a nonradial timelike geodesic, and let us denote
\begin{equation}\label{8.1}
E_J = 1 + {J_0}^2 / r^2,
\end{equation}
i.e. $E_J > 1$ is the $E$ for nonradial timelike geodesics.

{\bf Lemma 8.1}

A nonradial \textcolor[rgb]{0.00,0.00,1.00}{unbounded} timelike geodesic has its
turning point at larger $r$ than the radial one with the same $\Gamma$.

{\bf Proof}

The TP is where $\Gamma^2 - E \phi = 0$ in (\ref{3.3}). With $E = E_J$, its
equation is
\begin{equation}\label{8.2}
\left(\Gamma^2 / E_J - 1\right) r^2 + 2mr - e^2 = 0.
\end{equation}
\textcolor[rgb]{0.00,0.00,1.00}{When $\Gamma^2 / E_J < 1$, Eq. (\ref{8.2}) has
two solutions that obey
\begin{equation}\label{8.3}
r_{{\rm 1tp}\pm} = \frac 1 {1 - \Gamma^2/E_J(r_{\rm 1tp})}\ \left(m \pm
\sqrt{m^2 - e^2 + e^2 \Gamma^2/E_J(r_{\rm 1tp})}\right)
\end{equation}
with $r_{1{\rm tp}+} > r_+$ and $r_{1{\rm tp}-} < r_-$.} When $\Gamma^2 / E_J >
1$, the $r = r_{\rm 2tp}$ of the TP obeys
\begin{equation}\label{8.4}
r_{\rm 2tp} = \frac 1 {\Gamma^2/E_J(r_{\rm 2tp}) - 1}\ \left(-m + \sqrt{m^2 -
e^2 + e^2 \Gamma^2/E_J(r_{\rm 2tp})}\right)
\end{equation}
(the other solution of (\ref{8.2}) would have $r_{\rm 2tp} < 0$).

The case $\Gamma^2 / E_J = 1$ leads to $r_{\rm 3tp} = e^2 / (2m) < r_-$, but
this can happen only when $|J_0| = e^2 \sqrt{\Gamma^2 - 1} / (2m)$. The $r_{\rm
3tp}$ is larger than the $r_{\rm tp}$ of (\ref{3.10}), so Lemma 8.1 holds.

\textcolor[rgb]{0.00,0.00,1.00}{With $\Gamma^2 / E_J < 1$, the geodesics have
$r$ bounded ($\Gamma^2 - E \phi > 0$ for $r \in \left(r_{1{\rm tp}-}, r_{1{\rm
tp}+}\right)$), so the lemma does not apply.}

The $r_{\rm 2tp}$ of (\ref{8.4}) is smaller than $r_-$, which is rather easy to
verify. In Appendix \ref{r2tp} it is shown that it is larger than the $r_{\rm
tp}$ of (\ref{3.10}). $\square$

Thus, TPs for \textcolor[rgb]{0.00,0.00,1.00}{nonradial unbounded timelike
geodesics exist in the region $r \leq r_-$ when} $\Gamma^2 / E_J \geq 1$, which
translates to
\begin{equation}\label{8.5}
|J_0| \leq r \sqrt{\Gamma^2 - 1}.
\end{equation}
When $\Gamma^2 = 1$, this implies that $J_0 = 0$, i.e. a parabolic orbit can
enter the $r \leq r_-$ region only when it is radial. With $\Gamma^2
> 1$, Eq. (\ref{8.5}) is consistent with $r \leq r_-$ when
\begin{equation}\label{8.6}
|J_0| \leq \sqrt{\Gamma^2 - 1} r_-.
\end{equation}
Orbits with larger $|J_0|$ have TPs in $r > r_+$, i.e. outside the outer event
horizon.

With the parameters of (\ref{6.1}) -- (\ref{6.2}) the limit (\ref{8.6}) becomes
\begin{equation}\label{8.7}
|J_0| \leq 0.9 \times \sqrt{0.21} \approx 0.412431812546.
\end{equation}

We will now show (numerically) that even with $J_0 = 0.412$, which is close to
the maximum allowed value, the difference between the radial $G_1$ and the
nonradial timelike geodesic $G_J$ is insignificant from the point of view of our
main problem. The nonradial geodesics do not lie in any fixed $(u, v)$
coordinate plane, so for comparing $G_J$ with $G_1$ each point of $G_J$ will be
rotated (with unchanged $u$) into the $(u, v)$ plane of $G_1$.

\begin{figure}[h]
 ${}$ \\[-5cm]
 \hspace{-2cm} \includegraphics[scale=0.65]{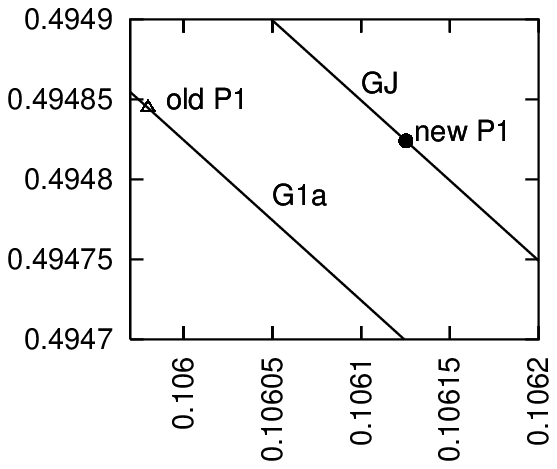}
 \hspace{-5mm} \includegraphics[scale=0.7]{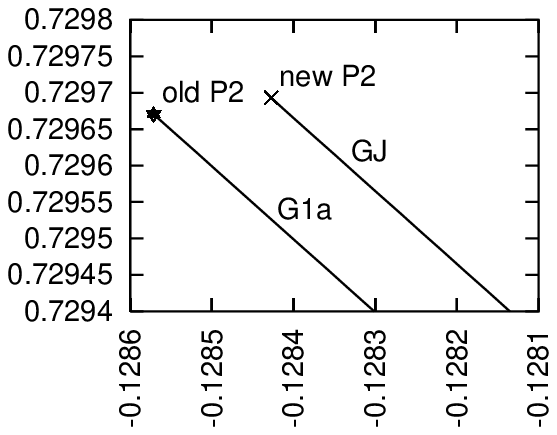}
 \caption{
 \label{geoJP1}
 \footnotesize
Comparison of the images of the radial $G_1$ and nonradial $G_J$ timelike
geodesics in the neighbourhood of points $P_1$ (left panel) and $P_2$ (right
panel) from Fig. \ref{drawgeo}. The two graphs are indistinguishable at the
scale of Fig. \ref{drawgeo}. See the text for more explanation.}
 \end{figure}

The $G_J$ goes off the same $(u_0, v_0)$ given by (\ref{6.4}) as G1a. In Fig.
\ref{drawgeo}, their images between E and the neighbourhood of $P_2$ are
indistinguishable. Figure \ref{geoJP1} shows their $({\cal U}, {\cal V})$
coordinates near $P_1$ (left panel) and $P_2$ (right panel, the scale in the two
panels is not the same). The `new $P_1$' is reached by $G_J$ at the same $r =
0.97$ as the old $P_1$ was reached by $G_1$. The `new $P_2$' is reached at the
same $r = 0.94$ as the old $P_2$.

\begin{figure}[h]
 \includegraphics[scale=0.5]{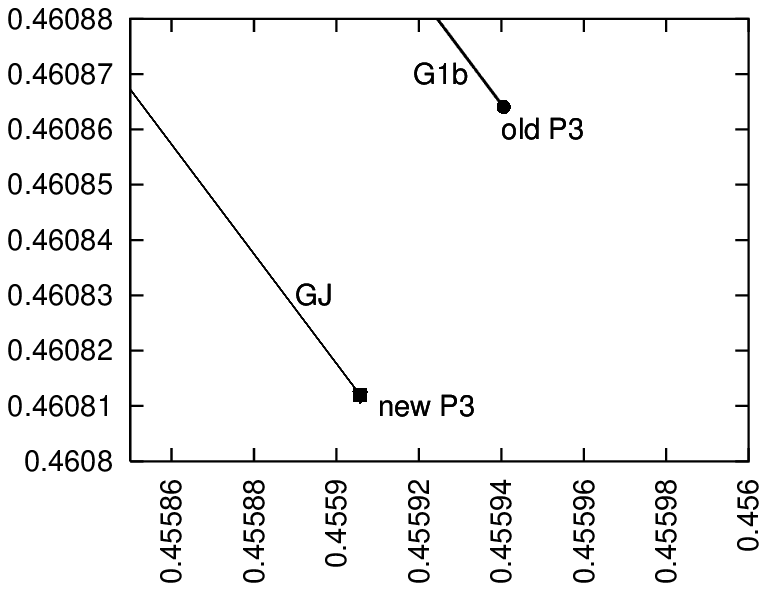}
 \hspace{-5mm} \includegraphics[scale=0.7]{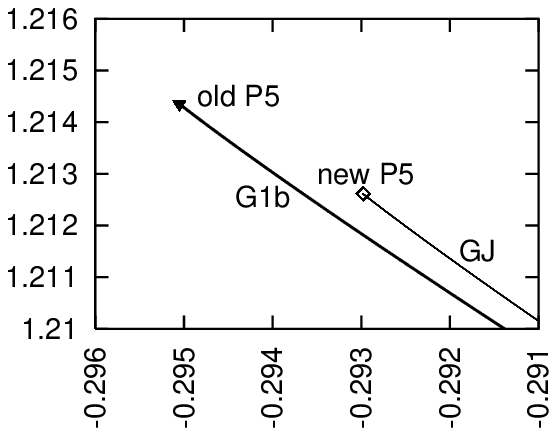}
 \caption{
 \label{geoJP5}
 \footnotesize
Comparison of the images of $G_1$ and $G_J$ in the neighbourhood of points $P_3$
(left panel) and $P_5$ (right panel) from Fig. \ref{drawgeo}. The scales in the
panels are different. See the text for more explanation.}
 \end{figure}

Also the image of the second segment of $G_J$ (from the neighbourhood of $P_3$
to the neighbourhood of $P_5$) coincides with that of G1b at the scale of Fig.
\ref{drawgeo}, and the images of their TPs are indistinguishable. The images of
these segments are seen as separate only after magnification -- see Fig.
\ref{geoJP5}. The left panel shows the image of the neighbourhood of their
initial points, the right panel shows the images of their final (turning)
points. The 'old $P_3$' is the image of the 'old $P_1$' from Fig. \ref{geoJP1}
under the transformation (\ref{4.6}) -- (\ref{4.7}), the 'new $P_3$' is the
image of the 'new $P_1$' under the same transformation. Incidentally, Figs.
\ref{geoJP1} and \ref{geoJP5} show that the transformation does not preserve the
ordering of events along the ${\cal U}$ coordinate axis: 'new $P_1$' is at a
larger ${\cal U}$ than 'old $P_1$', but 'new $P_3$' is at a smaller ${\cal U}$
than 'old $P_3$. The two panels of Fig. \ref{geoJP5} show that somewhere between
$P_3$ and $P_5$ the image of $G_J$ intersects G1b, so $G_J$ goes at a (slightly)
larger inclination to the ${\cal U}$ axis than G1b.

The $r$ of the TP of $G_J$ (i.e. the solution of (\ref{8.2})) can be determined
only numerically. With the parameters of given by (\ref{6.1}), (\ref{6.2}) and
$J_0 = 0.412$ it is
\begin{equation}\label{8.8}
r_{\rm Jtp} = 0.51034408730992153.
\end{equation}

\section{A nonradial null geodesic}\label{nunorad}

\setcounter{equation}{0}

The numerical calculation for this section uses the same algorithm as that of
Sec. \ref{norad}. The difference is that for a null geodesic $G_N$ we have
instead of (\ref{8.1})
\begin{equation}\label{9.1}
E = E_N = {J_0}^2 / r^2,
\end{equation}
so (\ref{8.2}) is replaced by the new equation of TPs
\begin{equation}\label{9.2}
\left(\Gamma^2 r^2 / {J_0}^2 - 1\right) r^2 + 2 m r - e^2 = 0.
\end{equation}
Equations (\ref{8.3}) and (\ref{8.4}) still apply with $E_J$ replaced by $E_N$,
and it is still true that with $\Gamma^2/E_N < 1$ we have $r_{\rm 1tp+} > r_+$.
With $\Gamma^2/E_N > 1$ it is still true that $r_{\rm 2tp} < r_-$. However, it
is not always true that $r_{\rm 2tp} > r_{\rm tp}$ for the $r_{\rm tp}$ of
(\ref{3.10}), because here $(1 - 1 / E_N)$ may have any sign depending on the
value of $r$, so (\ref{f.2}) does not imply (\ref{f.3}) in Appendix \ref{r2tp}.

In the first segment (from E to the neighbourhood of $P_2$) the image of $G_N$
does not differ from G1a of Fig. \ref{drawgeo} ($G_N$ runs between the same
values of $r$ as $G_1$). For completeness we note: the $({\cal U}, {\cal V})$
coordinates in the figure of the analogue of point $P_2$ in Fig. \ref{drawgeo}
are
\begin{equation}\label{9.3}
({\cal U}_2, {\cal U}_2)_{\rm null} = (-0.12938745931238965, \quad
0.72955345209300437),
\end{equation}
to be compared with (\ref{6.5}). The $({\cal U}, {\cal V})$ coordinates in the
figure for the analogue of point $P_3$ (i.e. at the start of the second segment)
are
\begin{equation}\label{9.4}
({\cal U}_3, {\cal V}_3)_{\rm null} = (0.45616174194929571, \quad
0.46112925805749150),
\end{equation}
to be compared with (\ref{6.6}). At the scale of Fig. \ref{drawgeo} the image of
the second segment of $G_N$ is again indistinguishable from G1b, and their
endpoints seem to coincide. The $r$ at the TP of the nonradial null geodesic is
\begin{equation}\label{9.5}
r_{\rm Ntp} = 0.43720631439290869,
\end{equation}
and it is reached by the second segment of $G_N$ with $({\cal U}, {\cal V})$ in
the figure being
\begin{equation}\label{9.6}
({\cal U}_5, {\cal V}_5)_{\rm null} = (-0.29636815953860662, \quad
1.2149002620359113),
\end{equation}
to be compared with (\ref{6.7}).

So, the conclusion reached for the timelike geodesic remains in power: with the
initial point given by (\ref{6.3}), the TP lies to the future of the past light
cone of E$'$.

\section{Geodesics that violate causality}\label{newsec}

\setcounter{equation}{0}

We will now present two more numerical examples of radial timelike geodesics
emitted in sector I, one of which clearly violates causality, and the other one
has its TP nearly on the past light cone of E$'$.

 \begin{figure}
 \begin{center}
 ${}$ \\[-1cm]
 \includegraphics[scale=0.7]{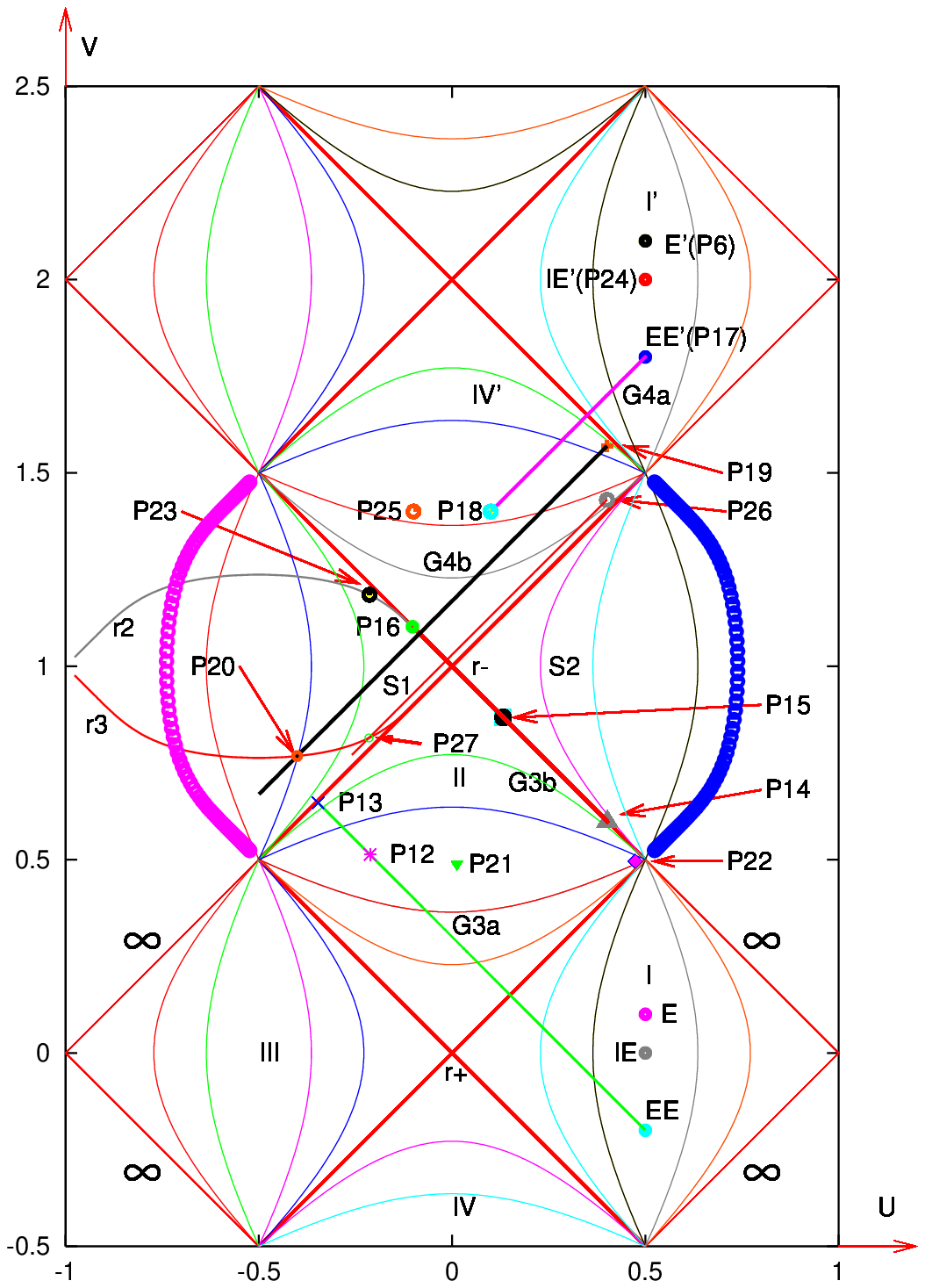}
 \caption{
 \label{drawlatear}
 \footnotesize
The radial timelike geodesic $G_3$ that violates causality and its corresponding
past-directed null geodesic $G_4$. The points IE and $P_{21}, P_{22}, P_{23}$
are characteristic points of the radial timelike geodesic that has its turning
point nearly on the past light cone (PLC) of IE$'$. The points $P_{24}, \dots,
P_{27}$ are characteristic points of the radial generator of the PLC of IE$'$.
See the text for detailed explanations. }
 \end{center}
 \end{figure}

For the first emission event we choose the point EE (see now Fig.
\ref{drawlatear}) whose $({\cal U}, {\cal V})$ coordinates (coinciding with the
internal $(U, V)$ in sector I) are
\begin{equation}\label{10.1}
({\cal U}, {\cal V}) = (0.5, -0.2),
\end{equation}
i.e. it lies below E at the same ${\cal U}$ in sector I. The ingoing radial
timelike geodesic emitted there, with $\Gamma$ and $E$ still given by
(\ref{6.2}), will be denoted $G_3$, and its first segment, denoted G3a, goes
from EE through $P_{12}$ to $P_{13}$ (at which, respectively, $r = 0.97$ and $r
= 0.94$, as before). The coordinates of $P_{12}$ and $P_{13}$ in the figure are
\begin{eqnarray}\label{10.2}
({\cal U}_{12}, {\cal V}_{12}) &=& (-0.21246353917544286, \quad
0.51294591609812235), \nonumber \\
({\cal U}_{13}, {\cal V}_{13}) &=& (-0.34654016129133364, \quad
0.64719713828109493).
\end{eqnarray}
The transformation (\ref{4.6}) -- (\ref{4.7}) is carried out at $P_{12}$ and
takes it to $P_{14}$. The second segment of $G_3$, denoted G3b, goes off
$P_{14}$ and continues to the intersection with the $r = r_{\rm tp}$ set, marked
$r_2$. The image of $P_{13}$ under the same transformation is $P_{15}$; this is
only a check of precision to show that $P_{15}$ lies on G3b. The $({\cal U},
{\cal V})$ coordinates in Fig. \ref{drawlatear} of $P_{14}$ and $P_{15}$ are
\begin{eqnarray}\label{10.3}
({\cal U}_{14}, {\cal V}_{14}) &=& (0.40181862662479906, \quad
0.597770308827682076 ), \nonumber \\
({\cal U}_{15}, {\cal V}_{15}) &=& (0.13195057342452676, \quad
0.86763568858480716).
\end{eqnarray}
The point of intersection of G3b with $r = r_2$, marked $P_{16}$ in Fig.
\ref{drawlatear}, has coordinates
\begin{equation}\label{10.4}
({\cal U}_{16}, {\cal V}_{16}) = (-0.10274640621741898, \quad
1.1023627992828624).
\end{equation}

Just like we did with the geodesic $G_1$, we now leave G3b behind for a while
and move to point EE$'$ (also denoted $P_{17}$) -- the first future copy of EE,
whose coordinates in the figure are $({\cal U}_{17}, {\cal V}_{17}) = (0.5,
1.8)$. From there, we issue the past-directed radial ingoing null geodesic
$G_4$, which is the radial generator of the past light cone of EE$'$. We denote
its first segment by G4a. As before, the equation of this geodesic can be
handled exactly. Using for $P_{17}$ the internal coordinates of sector I$'$ it
is
\begin{equation}\label{10.5}
V - U = -0.7.
\end{equation}
We continue this geodesic down to the point $P_{18}$ with the coordinates in the
figure
\begin{equation}\label{10.6}
({\cal U}_{18}, {\cal V}_{18}) = (0.1, 1.4),
\end{equation}
then transform its internal $(U, V) = (0.1, -0.6)$ to $(u, v)$ by (\ref{2.30})
-- (\ref{2.31}), then transform these to $(u', v')$ by (\ref{4.10}), then
transform $(u', v')$ to $(U', V')$ by (\ref{2.29}). What results is point
$P_{19}$ whose coordinates in the figure are
\begin{equation}\label{10.7}
({\cal U}_{19}, {\cal V}_{19}) = (0.400707810772957, \quad 1.56946086423524).
\end{equation}
{}From $P_{19}$ we now send the second segment of $G_4$, denoted G4b. Its
equation in the $({\cal U}, {\cal V})$ coordinates of Fig. \ref{drawlatear} is
\begin{equation}\label{10.8}
{\cal V} - {\cal U} = {\cal V}_{19} - {\cal U}_{19} = 1.168753053462283.
\end{equation}
The line G4b intersects $r_3$ (the locus of TPs of radial timelike geodesics) at
point $P_{20}$, which lies clearly later (has a smaller ${\cal U}$ coordinate)
than $P_{16}$. Consequently, the geodesic $G_3$ reaches its TP before crossing
the past light cone of EE$'$, so the observer at EE {\it can} send a message to
its own past and causality is broken. Here are the approximate $({\cal U}, {\cal
V})$ coordinates of $P_{20}$, read out from a magnification of its
neighbourhood:
\begin{equation}\label{10.9}
({\cal U}_{20}, {\cal V}_{20}) = (-0.401, 0.7678).
\end{equation}

Since a late-emitted geodesic had its TP later than (EE$'$)'s past light cone
(PLC), while the early-emitted one had its TP earlier than the PLC, there must
exist an intermediate emission point IE such that a radial timelike geodesic
$G_5$ going off there will have its TP right on the PLC. We will now construct a
numerical approximation to this geodesic. To avoid clogging Fig.
\ref{drawlatear} with too many details, we marked on it only the characteristic
points, without drawing the geodesic segments. Anticipating that IE must lie
close to the point of time-symmetry of sector I, we choose its coordinates at
\begin{equation}\label{10.10}
({\cal U}, {\cal V}) = (0.5, 0.0).
\end{equation}
On $G_5$, we choose the point $P_{21}$ at the same $r = 0.97$ as on the
previously constructed geodesics. Its coordinates in the figure are
\begin{equation}\label{10.11}
({\cal U}_{21}, {\cal V}_{21}) = (0.012375519215587246, \quad
0.48841064854212529).
\end{equation}
At $P_{21}$ we apply the transformation (\ref{4.6}) -- (\ref{4.7}) that takes it
to point $P_{22}$ of coordinates
\begin{equation}\label{10.12}
({\cal U}_{22}, {\cal V}_{22}) = (0.47402568470461826, \quad
-0.0041480326881506938).
\end{equation}
The second segment of $G_5$ goes off $P_{22}$ and intersects the locus of TPs at
$P_{23}$ with
\begin{equation}\label{10.13}
({\cal U}_{23}, {\cal V}_{23}) = (-0.21394900777138914, \quad
1.1848938941310596).
\end{equation}

\begin{figure}[h]
 \begin{center}
 ${}$ \\[-3.5cm]
 \includegraphics[scale=0.65]{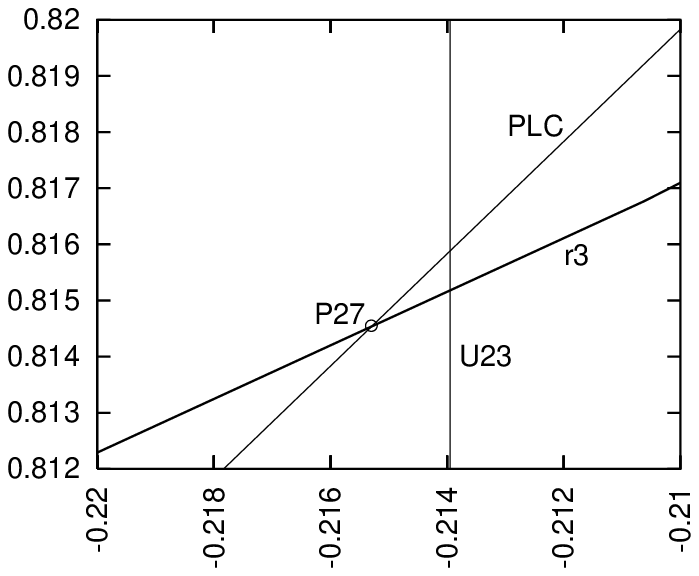}
 \caption{
 \label{drawlatearlupa2}
 \footnotesize
A closeup view on the neighbourhood of point $P_{27}$ in Fig. \ref{drawlatear}.
The line U23 marks the ${\cal U}$ coordinate of point $P_{23}$. Ideally, it
should go through $P_{27}$, which is the point of intersection of the past light
cone (PLC) of point IE$'$ with $r_3$. It can be made arbitrarily near $P_{27}$
by slightly increasing the coordinae ${\cal V}$ of IE (and, correspondingly, of
IE$'$). }
 \end{center}
 \end{figure}

Now we proceed to the point IE$'$, also denoted $P_{24}$, which is the first
future copy of IE. Its coordinates in the figure are
\begin{equation}\label{10.14}
({\cal U}_{24}, {\cal V}_{24}) = (0.5, 2.0).
\end{equation}
{}From there, we issue the past-directed null geodesic $G_6$ (the radial
generator of the PLC of IE$'$) and follow it to the point $P_{25}$ of
coordinates
\begin{equation}\label{10.15}
({\cal U}_{25}, {\cal V}_{25}) = (-0.1, 1.4).
\end{equation}
To $P_{25}$ we apply the transformation (\ref{4.10}) that takes it to $P_{26}$
of coordinates
\begin{equation}\label{10.16}
({\cal U}_{26}, {\cal V}_{26}) = (0.400707810772957, \quad 1.43053913576476).
\end{equation}
{}From $P_{26}$ we send the second segment of $G_6$ and continue it slightly
beyond the intersection with $r_3$, denoted $P_{27}$. It has the approximate
(read out from the figure) coordinates
\begin{equation}\label{10.17}
({\cal U}_{27}, {\cal V}_{27}) \approx (-0.2153, 0.81455).
\end{equation}
The ${\cal U}_{27}$ should coincide with ${\cal U}_{23}$, and it does so up to
0.00135, which we will be satisfied to accept as the proof of our thesis. Figure
\ref{drawlatearlupa2} shows this discrepancy.

\section{Conclusions and summary}\label{conclu}

\setcounter{equation}{0}

The main purpose of this paper was to verify whether an observer in an
asymptotically flat region of the maximally extended Reissner -- Nordstr\"{o}m
(RN) spacetime with $e^2 < m^2$ can send a message (by means of timelike or null
geodesics) to the past of his/her first future copy (see Figs. \ref{reinormax},
\ref{drawgeo} and \ref{drawlatear}). If this is possible, then the
identifications of the asymptotically flat regions do lead to acausality. The
answer is that geodesics emitted sufficiently late do not break causality, while
those emitted sufficiently early do. Let the initial point of the geodesic be at
event E, and let the first future copy of E be E$'$. Then any radial null
geodesic emitted at E will hit the singularity and stop there. For late-emitted
timelike geodesics the turning point (TP) in the tunnel between the
singularities lies to the future of the past light cone (PLC) of E$'$, so the
signal may come back to the observer only later than it was emitted. But for
early-emitted timelike geodesics, the opposite happens: the TP lies to the past
of the PLC of E$'$. In this case, the geodesic continued beyond the TP can cross
the worldline of E$'$ in the past of E$'$, so the breach of causality is
possible. The border between 'early' and 'late' are those emission points for
which the TP lies right on the PLC of E$'$.

Here is a step-by-step summary of this paper.

After presenting the main idea and the overview of this paper in Sec.
\ref{intro}, the geometric properties of the Reissner -- Nordstr\"{o}m (RN)
spacetime were discussed in Sec. \ref{RNintro}. For the needs of the further
sections, this discussion, in particular the derivation of the maximal
extension, was done in more detail than in the original paper \cite{GrBr1960}.
Several interesting facts emerged that may evade the attention of a reader of
the basic sources (and perhaps evaded the attention of the authors of Refs.
\cite{GrBr1960} and \cite{Cart1973}); see below for more on this.

Section \ref{geode} contains a discussion of the geodesic equations in the $(u,
v)$ coordinates that remove the spurious singularities at the horizons. Also
here, a few interesting facts emerged, see below.

Section \ref{thepro} contains the following results:
\begin{enumerate}
\item Some additional relations between the $(u, v)$ coordinates (see above)
and the $(U, V)$ coordinates of the conformal image of the RN spacetime, used in
the figures.

\item The transformations between the $(u, v)$ coordinates removing the
singularity at $r = r_+$ and the $(u', v')$ that remove the singularity at $r =
r_-$. It turned out that this transformation leaves the geodesic equation
unchanged.
\end{enumerate}

In Sec. \ref{shaper}, the transformation of the $(U, V)$ coordinates needed to
self-consistently place in one figure the various sectors of the maximal
extension was identified and illustrated using three figures.

In Secs. \ref{numcal} and \ref{PLCE}, an exemplary radial timelike geodesic
$G_1$ going off event E in an asymptotically flat region of RN into the black
hole was numerically calculated. Such a geodesic cannot hit the true singularity
at $r = 0$ \cite{GrBr1960,PlKr2024}, but has a TP at $r = r_{\rm tp} \in (0,
r_-)$. The question was whether, after bouncing at $r_{\rm tp}$, it can enter
the PLC of E$'$ -- the first future copy of E. The answer, for this particular
geodesic, turned out to be 'no': the radial generator $G_2$ of the PLC of E$'$
can be calculated exactly, and the TP of $G_1$ lies to the future of $G_2$.

In Sec. \ref{norad}, the consideration of Sec. \ref{numcal} was applied to a
nonradial timelike geodesic $G_J$ going off the same event E with the angular
momentum constant $J_0$ having a nearly-maximum absolute value. (With larger
$|J_0|$, timelike geodesics do not enter the tunnel between the singularities
and are irrelevant for the problem of causality.) The result was the same as in
Sec. \ref{numcal}: the turning point lies to the future of the PLC of E$'$.

In Sec. \ref{nunorad}, still the same consideration was applied to a nonradial
null geodesic $G_N$ going off the same E with the same $|J_0|$ as in Sec.
\ref{norad}. The result was still the same: the TP of $G_N$ lies to the future
of the PLC of E$'$.

In Sec. \ref{newsec}, the construction of Sec. \ref{numcal} was applied to two
other emission events, EE and IE, having the same spatial coordinate $V$ as E.
The event EE was earlier than E while IE was such that the radial timelike
geodesic emitted there has its TP nearly on the PLC of IE$'$. For EE, the break
of causality does occur: the TP of the timelike geodesic emitted there lies to
the past of the PLC of EE$'$. The second calculation proved that the logical
conclusion from the earlier calculations does indeed hold: such an IE exists.

In addition to these results, the paper revealed a few properties of the
maximally extended RN spacetime that remain unnoticed when one reads general
descriptions of the maximal extension procedure, such as in Refs.
\cite{GrBr1960} and \cite{PlKr2024}. Here is the recapitulation:

\begin{enumerate}
\item The transformation given in Ref. \cite{GrBr1960} from the curvature
coordinates $(t, r)$ to the $(u, v)$ coordinates that removes one or the other
spurious singularity is valid on one side of the horizon. To cover the whole
range of $(u, v)$ one needs to consider four cases, corresponding to different
signs of $u$, $v$ and $u^2 - v^2$, see Sec. \ref{RNintro} and Fig.
\ref{sectors}.

\item The shapes of the $r =$ constant lines in the $(U, V)$ coordinates depend
on the values of the $m$ and $e$ constants, see Fig. \ref{rconst} (in the $(u,
v)$ coordinates they are always hyperbolae, variously placed). Papers and books
published so far used a schematic illustrative representation of these lines
that is correct only for sufficiently small values of the distance between the
foci of the hyperbola $u^2 - v^2 =$ constant.

\item The geodesic equations in the $(u, v)$ coordinates are the same on both
sides of the horizon $r = r_+$, see the comment under (\ref{3.7}).

\item The geodesic equations in the $(u, v)$ coordinates are invariant under the
transformation $(u, v) = (v', u')$, see (\ref{3.8}) -- (\ref{3.9}).

\item The geodesic equation is not changed by the transformation between the
$(u, v)$ coordinates that remove the singularity at $r = r_+$ and those that
remove the singularity at $r = r_-$ (Eqs. (\ref{4.3}) -- (\ref{4.7})), compare
(\ref{3.11}) with (\ref{4.8}).

\item Matching the conformal images of the regions $r > r_-$ and $r < r_+$
requires reflecting the latter in the line $V = -U + 0.5$. The required
transformation of $(U, V)$ is simple, see the last paragraph of Sec.
\ref{shaper}.
\end{enumerate}

In brief, the main conclusion of this paper is that identifications of
asymptotically flat regions of the maximally extended RN spacetime do lead to
acausality (the possibility of sending messages to one's own past, or, in other
words, the existence of closed timelike lines). As an additional bonus, the
readers now have a clear recipe to follow when calculating geodesics propagating
from one $(U, V)$ map to another.
\appendix

\section{Finding $r$ from a given ${\rm e}^{2 \gamma r^*}$ using
(\ref{2.21})}\label{anal219}

\setcounter{equation}{0}

In the range $r > r_-$ and with $\gamma = \gamma_1$, Eq. (\ref{2.21}) becomes
\begin{equation}\label{a.1}
{\rm e}^{2 \gamma_1 r^*} = {\rm e}^{2 \gamma_1 r} \left|r - r_+\right| \left(r -
r_-\right)^{-(r_- / r_+)^2}.
\end{equation}
This decreases monotonically from $\infty$ at $r \to r_-$ to zero at $r = r_+$,
then increases monotonically from 0 to $+ \infty$ at $r \to \infty$. Thus, given
${\rm e}^{2 \gamma_1 r^*}$, the algorithms for finding $r > r_-$ by the
bisection method are different for $r < r_+$ and $r > r_+$.

In the range $0 < r < r_+$, and with $\gamma = \gamma_2$, Eq. (\ref{2.21})
implies the following. For $0 < r <r_-$ the function ${\rm e}^{2 \gamma_2 r^*}$
decreases from ${r_+}^{- (r_+ / r_-)^2} r_-
> 0$ at $r = 0$ to zero at $r \to r_-$. For $r_- < r < r_+$, it increases from
zero at $r \to r_-$ to $+ \infty$ at $r = r_+$. Consequently, also in this case
the algorithms for finding $r$ are different for $0 < r < r_-$ and for $r_- < r
< r_+$.

\section{Crossing $r = r_{\pm}$ with Eq. (\ref{3.11})}\label{numer}

\setcounter{equation}{0}

To avoid encountering an expression of the type 0/0 while crossing $r = r_{\pm}$
with $\sigma = -1$, we apply the rule given below to the denominator in
(\ref{3.11}):
\begin{equation}\label{b.1}
x - y = \frac {x^2 - y^2} {x + y}.
\end{equation}
Using (\ref{2.15}) to eliminate $u^2 - v^2$ in two places, we obtain
\begin{equation}\label{b.2}
\dr u v = \frac {- \Gamma \sqrt{\Gamma^2 - E \phi} + E \gamma^2 u v f^2}
{\Gamma^2 + E (\gamma v f)^2}.
\end{equation}
Since $E \geq 0$, (\ref{b.2}) is well-defined also where $\phi < 0$. In the
limit ($\phi \to 0, v \to \pm u$), Eq. (\ref{b.2}) becomes (because of $\Gamma >
0$)
\begin{equation}\label{b.3}
\lim_{\phi \to 0, v \to \pm u} \left(\dr u v\right) = \frac {- \Gamma^2 \pm E
(\gamma u f)^2} {\Gamma^2 + E (\gamma u f)^2}.
\end{equation}
With the minus sign this becomes $\dril u v = -1$.

To deal with the 0/0 problem for (\ref{3.11}) with $\sigma = +1$, we apply $x +
y = (x^2 - y^2) / (x - y)$ instead of (\ref{b.1}) and again use (\ref{2.15}) to
eliminate $u^2 - v^2$. The result is
\begin{equation}\label{b.4}
\dr u v = \frac {\Gamma \sqrt{\Gamma^2 - E \phi} + E \gamma^2 u v f^2} {\Gamma^2
+ E (\gamma v f)^2}.
\end{equation}

\section{Calculating $r$ from a given $r^*$ by the bisection method using
(\ref{2.9})}\label{bisec}

\setcounter{equation}{0}

The equation to be solved for $r$ is
\begin{equation}\label{c.1}
G(r) \df {\rm e}^{2 \gamma_1 r^*} = {\rm e}^{2 \gamma_1 r} \left|r - r_+\right|
\left(r - r_-\right)^{- a}, \qquad a \df {r_-}^2 / {r_+}^2,
\end{equation}
where the value of $G(r)$ is given (the second pair of the $||$ brackets in
(\ref{2.21}) was replaced with $()$ because (\ref{c.1}) will be applied only
where $r > r_-$).\footnote{With $\gamma = \gamma_2$, the corresponding
$\widetilde{G}(r)$ is finite for all $r \in (0, r_+)$. The choice of the initial
bounding values for $r$ is $(r_-, r_+)$ in the segment where $r_- < r < r_+$ and
$(0, r_-)$ in the segment where $0 < r < r_-$.}

\begin{figure}[h]
 \begin{center}
 \includegraphics[scale=0.6]{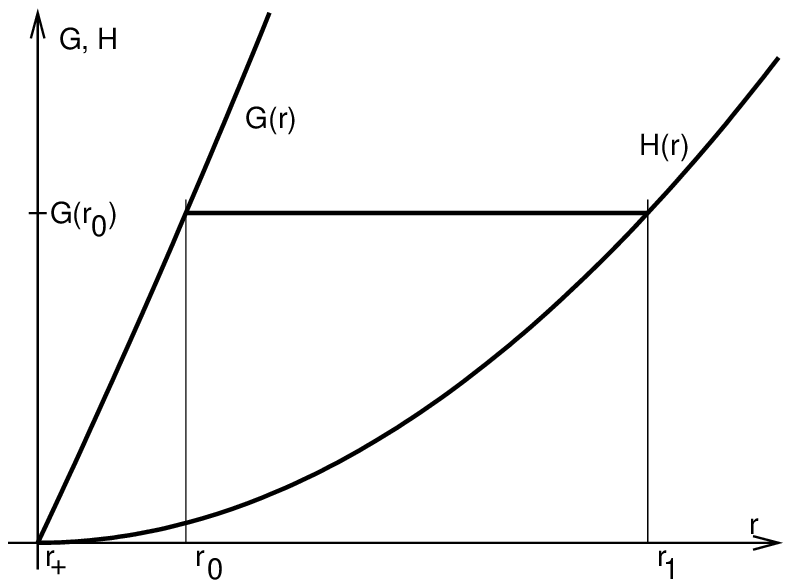}
 \caption{
 \label{major}
 \footnotesize
In finding $r_0$ from a given $G(r_0)$ by the bisection method, the initial
upper bound for $r_0$ is $r_1$. See the text for more explanation.}
 \end{center}
 \end{figure}

In the first segments of the geodesics $G_1$, $G_3$ and $G_5$ in Secs.
\ref{numcal} and \ref{newsec}, the initial bounding values for $r$ are $r = r_-$
and $r = r_0$, where $r_0$ is any $r$ larger than the $r$ coordinate of the
initial point E. But $r(E)$ can be anywhere in $(r_+, +\infty)$. Therefore, the
upper bound for $r_0$ is not self-evident. The method of determining it is
illustrated in Fig. \ref{major}, and here is the description.

The function $G(r)$ is monotonic for $r > r_+$ by (\ref{2.7}) because $\dril
{r^*} r = 1 / \phi > 0$ in this range, and it varies between 0 and $+\infty$. We
need to find such a function $H(r)$ that also varies between 0 and $+\infty$ and
$G(r) > H(r)$ in $(r_+, +\infty)$. Then, for any $r_0$ there exists an $r_1 >
r_0$ such that $G(r_0) = H(r_1)$. This $r_1$ is an initial upper bound for
$r_0$. The problem is to ensure that the equation $G(r_0) = H(r_1)$ can be
explicitly solved for $r_1$.

Remembering that the problem arises in the region $r > r_+ > r_-$, we have
\begin{equation}\label{c.2}
0 < \frac {r - r_+} {r - r_-} < 1 \quad \Longrightarrow \quad \left(\frac {r -
r_+} {r - r_-}\right)^c > \frac {r - r_+} {r - r_-} \quad {\rm when}\ c < 1
\Longrightarrow
\end{equation}
\begin{equation}\label{c.3}
G(r) \equiv {\rm e}^{2 \gamma_1 r} \left(r - r_+\right)^{1 - a} \left(\frac {r -
r_+} {r - r_-}\right)^a > {\rm e}^{2 \gamma_1 r} \left(r - r_+\right)^{2 - a} /
\left(r - r_-\right)
\end{equation}
by virtue of (\ref{c.2}) because $a = {r_-}^2/{r_+}^2 < 1$. Moreover
\begin{equation}\label{c.4}
{\rm e}^{2 \gamma_1 r} > {\rm e}^{2 \gamma_1 (r - r_-)} > 2 \gamma_1 \left(r -
r_-\right),
\end{equation}
since $\gamma_1 > 0$ and ${\rm e}^x > x$ for all $x \in (- \infty, + \infty)$.
Thus finally
\begin{equation}\label{c.5}
G(r) > 2 \gamma_1 \left(r - r_+\right)^{2 - a} \df H(r)
\end{equation}
and the upper bound for $r_0$ is
\begin{equation}\label{c.6}
r_1 = r_+ + \left[\frac {G(r_0)} {2 \gamma_1}\right]^{1 / (2 - a)}.
\end{equation}

\section{Proof of Lemma 4.1}\label{dVbydv}

\setcounter{equation}{0}

Along a line of constant $r$ we have $u = \sqrt{C + v^2}$ from (\ref{2.13}) --
(\ref{2.19}), so from (\ref{2.29})
\begin{equation}\label{d.1}
\dr V v = 2\  \frac {1 + \cosh(2v) \cosh(2u) - (v / \sqrt{C + v^2}) \sinh(2v)
\sinh(2u)} {[\cosh(2u) + \cosh(2v)]^2}.
\end{equation}
Since $\sinh x < \cosh x$ for all finite $x$, we have $\sinh(2v) \sinh(2u) <
\cosh(2v) \cosh(2u)$. With $C > 0$, $v / \sqrt{C + v^2} < 1$, so $\dril V v >
0$. $\square$

With $C < 0$, Lemma 4.1 may hold or not, depending on the value of $|C|$ and the
range of $v$.

\section{Calculating $\dril {u'} {v'}$ from (\ref{4.4}) --
(\ref{4.7})}\label{dubydv}

\setcounter{equation}{0}

We use the identity
\begin{equation}\label{e.1}
\dr u v = \frac {\dril u {v'}} {\dril v {v'}} \equiv \frac {\pdril u {v'} +
(\pdril u {u'}) (\dril {u'} {v'})} {\pdril v {v'} + (\pdril v {u'}) (\dril {u'}
{v'})}
\end{equation}
and equate this to (\ref{3.11}). We solve the resulting equation for $\dril {u'}
{v'}$, then substitute for $u$ and $v$ from (\ref{4.4}) -- (\ref{4.5}). The
final result is (\ref{4.8}), after quite some algebra.

\section{A nonradial timelike geodesic going through the tunnel between the
singularities has its turning point at a larger $r$ than the radial one}
\label{r2tp}

\setcounter {equation} {0}

We verify this statement by contradiction. Let
\begin{equation}\label{f.1}
r_{\rm 2tp} \leq r_{\rm tp},
\end{equation}
where $r_{\rm 2tp}$ is given by (\ref{8.4}) and $r_{\rm tp}$ by (\ref{3.10}).
Since $\Gamma^2 - 1 > 0$ and $\Gamma^2/E_J - 1 > 0$, the resulting inequality
may be written as
\begin{eqnarray}\label{f.2}
&& \left(\Gamma^2 - 1\right)\ \left(- m + \sqrt{m^2 - e^2 + e^2
\Gamma^2/E_J(r_{\rm 2tp})}\right) \nonumber \\
&& \leq \left(\Gamma^2/E_J(r_{\rm 2tp}) - 1\right)\ \left(- m +\sqrt{m^2 - e^2 +
e^2 \Gamma^2}\right).
\end{eqnarray}
Both sides above are positive, so they may be squared and the direction of the
inequality will not change. After squaring, using (\ref{f.2}) to eliminate \\
$\left(\Gamma^2 - 1\right)\ \sqrt{m^2 - e^2 + e^2 \Gamma^2/E_J(r_{\rm 2tp})}$
and simplifying, $\left(\Gamma^2/E_J - 1\right) \left(1 - 1 / E_J\right) > 0$
factors out, and what remains may be written as
\begin{equation}\label{f.3}
2m \Gamma^2\ \sqrt{m^2 - e^2 + e^2 \Gamma^2} \geq \Gamma^2 \left(2m^2 - e^2 +
e^2 \Gamma^2\right) > 0.
\end{equation}
After both sides of this are squared and the result is simplified, $e^4
\left(\Gamma^2 - 1\right)^2 \leq 0$ follows. This can hold only when $e = 0$ or
$\Gamma^2 = 1$. The first case is the Schwarzschild limit, the second one is
dealt with in the main text below (\ref{8.5}). So, in generic cases, $r_{\rm
2tp}
> r_{\rm tp}$. $\square$
\bigskip

{\bf Acknowledgement.} For some calculations, the computer algebra system
Ortocartan \cite{Kras2001,KrPe2000} was used.

\end{document}